\PassOptionsToPackage{pdfpagelabels=false}{hyperref}   
\expandafter\def\csname ver@fixltx2e.sty\endcsname{}   

\documentclass[fleqn,usenatbib]{mnras}

\usepackage{graphicx}	
\usepackage{amsmath}	
\usepackage{amssymb}	
\usepackage{siunitx}
\usepackage{makecell}

\usepackage{txfonts}

\usepackage[T1]{fontenc}
\usepackage{ae,aecompl}

\title{SDSS-IV MaNGA: spatially resolved dust attenuation in spiral galaxies}

\author[M. J. Greener et al.]{
Michael J. Greener,$^{1}$\thanks{E-mail: michael.greener@nottingham.ac.uk}
Alfonso Arag{\'o}n-Salamanca,$^{1}$
Michael R. Merrifield,$^{1}$
\newauthor
Thomas G. Peterken,$^{1}$
Amelia Fraser-McKelvie,$^{1}$
Karen L. Masters,$^{2}$
\newauthor
Coleman M. Krawczyk,$^{3}$
Nicholas F. Boardman,$^{4}$
M{\'e}d{\'e}ric Boquien,$^{5}$
\newauthor
Brett H. Andrews,$^{6}$
Jonathan Brinkmann,$^{7}$
and Niv Drory$^{8}$
\\
$^{1}$School of Physics \& Astronomy, University of Nottingham, University Park, Nottingham, NG7 2RD, UK\\
$^{2}$Haverford College, Department of Physics and Astronomy, 370 Lancaster Avenue, Haverford, Pennsylvania 19041, USA\\
$^{3}$Institute of Cosmology \& Gravitation, University of Portsmouth, Dennis Sciama Building, Portsmouth, PO1 3FX, UK\\
$^{4}$Department of Physics \& Astronomy, University of Utah, Salt Lake City, Utah 84112, USA\\
$^{5}$Centro de Astronom{\'i}a, Universidad de Antofagasta, Avenida Angamos 601, Antofagasta 1270300, Chile\\
$^{6}$Department of Physics and Astronomy, University of Pittsburgh, 3941 O'Hara Street, Pittsburgh, Pennsylvania 15260, USA\\
$^{7}$Apache Point Observatory, P.O. Box 59, Sunspot, New Mexico 88349, USA\\
$^{8}$McDonald Observatory, The University of Texas at Austin, 1 University Station, Austin, Texas 78712, USA\\
}

\date{Accepted XXX. Received YYY; in original form ZZZ}

\pubyear{2020}

\begin{document}
\label{firstpage}
\pagerange{\pageref{firstpage}--\pageref{lastpage}}
\maketitle

\begin{abstract}
Dust attenuation in star-forming spiral galaxies affects stars and gas in different ways due to local variations in dust geometry. We present spatially resolved measurements of dust attenuation for a sample of 232 such star-forming spiral galaxies, derived from spectra acquired by the SDSS-IV MaNGA survey. The dust attenuation affecting the stellar populations of these galaxies (obtained using full spectrum stellar population fitting methods) is compared with the dust attenuation in the gas (derived from the Balmer decrement). Both of these attenuation measures increase for local regions of galaxies with higher star formation rates; the dust attenuation affecting the stellar populations increases more so than the dust attenuation in the gas, causing the ratio of the dust attenuation affecting the stellar populations to the dust attenuation in the gas to decrease for local regions of galaxies with higher star formation rate densities. No systematic difference is discernible in any of these dust attenuation quantities between the spiral arm and inter-arm regions of the galaxies. While both the dust attenuation in the gas and the dust attenuation affecting the stellar populations decrease with galactocentric radius, the ratio of the two quantities does not vary with radius. This ratio does, however, decrease systematically as the stellar mass of the galaxy increases. Analysis of the radial profiles of the two dust attenuation measures suggests that there is a disproportionately high concentration of birth clouds (incorporating gas, young stars and clumpy dust) nearer to the centres of star-forming spiral galaxies.
\end{abstract}




\section{Introduction}
\label{sec:Introduction}

Properties of galaxies such as gas metallicities and star formation rates (SFRs) are derived from nebular emission lines produced as a result of gas heated by star-forming (SF) regions. However, this light at optical and ultraviolet (UV) wavelengths is absorbed and scattered by dust between the emission source and the observer, which results in dust attenuation.\footnote{\emph{Extinction} is defined as absorption + scattering by dust grains out of the observer's line of sight. This phenomenon is not to be confused with \emph{attenuation}, which is absorption + scattering into and out of the observer's line of sight. These terms are often erroneously used interchangeably -- see Fig. 1 of \citet{Salim2020TheGalaxies} for further clarification.} We refer the reader to \citet{Kewley2019UnderstandingLines} and \citet{Salim2020TheGalaxies} for recent reviews on emission lines and dust attenuation respectively, and \citet{Calzetti2001TheGalaxies} for a pioneering study on dust within star-forming galaxies. SF spiral galaxies are undergoing dynamical and structural evolution (see reviews in \citealp{Boselli2006EnvironmentalClusters} and \citealp{Sanchez2019Spatially-ResolvedGalaxies}), and are often associated with regions with considerable SFRs \citep[e.g.][]{Roberts1969Large-ScaleFormation, Kennicutt1998TheGalaxies, Boselli2006EnvironmentalClusters, Sanchez2019Spatially-ResolvedGalaxies}. These findings are of particular interest for this work, as SFRs are often correlated with the total dust content of SF galaxies \citep[e.g.][]{DaCunha2008, daCunha2010NewGalaxies, Rowlands2014Herschel-ATLAS:Redshifts, Remy-Ruyer2015LinkingPicture}.

Models of dust attenuation usually consider two main components in spiral galaxies: (1) \emph{an interstellar medium (ISM)}, which is diffuse and increases in density closer to the galactic centre \citep[e.g.][]{Peletier1995TheGalaxies, Boissier2004TheGalaxies, Munoz-Mateos2009RADIALPROPERTIES, Wild2011a, GonzalezDelgado2015TheSequence, Goddard2016SDSS-IVType}; and (2) \emph{birth clouds (BCs)}, in which dust is very highly concentrated, clumpy, and therefore optically thicker than the ISM \citep[e.g.][]{LonsdalePersson1987OnFluxes, Charlot2000AGalaxies, Calzetti2001TheGalaxies, Wild2011a, Wuyts2013A1.5, Price2014DirectRates}. While all stars are born within these dense clouds, stellar migration means that only the youngest, hottest and most massive O and B stars remain in close proximity to their BCs \citep[e.g.][]{Walborn1999SomeHST/NICMOS, Sellwood2002RadialDiscs, Roskar2008RidingDisks, Minchev2009AOverlap, Loebman2010TheMigration, Frankel2018MeasuringDisk, Minchev2018EstimatingGradient, Feltzing2020ConstrainingStudy}. These massive, young stars then ionise the surrounding gas, forming \textsc{H~ii} regions within the BCs. The central SF region is surrounded by clumps of gas and dust -- propelled outwards by outflows from the starburst site -- which absorb both continuum photons and those from the \textsc{H~ii} regions alike \citep{Charlot2000AGalaxies, Calzetti2001TheGalaxies}.

We ultimately want to better understand the nature of dust attenuation, and how it affects astronomical observations. The motivation for doing so is compelling. The dust content of galaxies has been shown to correlate with their gas metallicities and gas contents \citep{Calzetti2001TheGalaxies}, but its presence reduces and reddens both the UV and optical flux emerging from host galaxies. This effect results in biased measurements of SFRs \citep{Madau1996, Steidel1999LymanRedshift, Glazebrook1999, Sullivan2000AnSample, Bell2001ARates, daCunha2010NewGalaxies}, SFRs per unit physical surface area \citep{Buat1989StarGalaxies., Kennicutt2007StarLaw, Li2019InterpretingMaNGA}, SFR densities \citep{Brinchmann2004TheUniverse}, specific SFRs \citep{Buat2006TheFormation, Pannella2009StarDownsizing}, gas metallicities \citep{Gilbank2010TheLuminosities, Wilkinson2015a}, mass and luminosity densities \citep{Brinchmann2004TheUniverse, Popescu2007}, and even distance estimates \citep{Giovanelli1995DependenceExtinction}.

\medskip

Previous authors \citep[e.g.][]{Fanelli1988SpectralGalaxies, Storchi-Bergmann1994UltravioletEffects, Calzetti1994DustLaw, Calzetti2000TheGalaxies, Kreckel2013a} have extensively quantified and compared two independent measures of dust attenuation. Firstly, measurements of the the ratio of the Balmer emission lines $\rm H \alpha$ to $\rm H \beta$ are used as a diagnostic to derive the dust attenuation in the gas. The ratio of $\rm H \alpha$ to $\rm H \beta$ is known as the Balmer decrement. Since both lines are produced by the recombination of gas ionised by young stars, the Balmer decrement provides a measure of the dust attenuation due to the clumpy BCs, and so is useful to map the distribution of these dust clumps within the SF galaxies. The second measure is of the dust attenuation affecting the stellar populations. The gas is associated with the young stars close to the plane of the host galaxies, where dust is concentrated; the stars, by contrast, have a much greater scaleheight on average. Furthermore, since the dust and gas are well-mixed, we therefore might expect the dust attenuation in the gas to be higher than that affecting the stellar populations. Previous work, in which the ratio of these two dust attenuation measures has been calculated, has determined that this is indeed the case \citep[e.g.][]{Fanelli1988SpectralGalaxies, Storchi-Bergmann1994UltravioletEffects, Calzetti1994DustLaw, Calzetti2000TheGalaxies, Kreckel2013a}.

The measurements made by many of these previous authors, are, however, limited by the lack of spatial resolution. Fortunately, integral field spectroscopic surveys, such as the Calar Alto Legacy Integral Field Area \citep[CALIFA;][]{Sanchez2012CALIFASurvey} survey, the Multi Unit Spectroscopic Explorer Wide \citep[MUSE-Wide;][]{Urrutia2019TheRelease} survey, and the largest such survey, Mapping Nearby Galaxies at Apache Point Observatory \citep[MaNGA;][]{Bundy2015OVERVIEWOBSERVATORY}, now allow for spatially resolved observations of nearby spiral galaxies with a representative range of masses and SFRs. Such observations provide spatially resolved measurements for each of the dust attenuation derived from the gas, the dust attenuation affecting the stellar populations, and the excess between the former quantity and the latter. This final metric is perhaps a more physically meaningful property than the ratio between the two dust attenuation measures determined by previous studies, since it quantifies the excess dust attenuation close to the plane of the galaxy.

\medskip

In this work, we quantify the dust attenuation in a well-defined sample of 232 nearby SF spiral galaxies with the aid of integral field spectroscopy. Two independent, but complementary, methods are used: first by using the Balmer decrement as a diagnostic for the dust attenuation in the gas, and second by using full-spectrum stellar population fitting methods to measure the dust attenuation affecting the stellar populations in the same galaxies.

We will use this analysis to propose a new geometry for the dust distribution in SF spiral galaxies, as well as putting better constraints on the difference between the optical depth of the Balmer emission lines $\rm H \alpha$ and $\rm H \beta$ and that affecting the stellar continuum. It is relatively easy to use measurements of the Balmer decrement to determine the dust attenuation in low-redshift galaxies \citep[e.g.][]{Sullivan2000AnSample, Bell2001ARates, Kreckel2013a} and in high-redshift galaxies \citep[e.g.][]{Glazebrook1999}. For low-redshift galaxies, while the attenuation affecting the stellar populations can also be readily measured \citep[e.g.][]{Sullivan2000AnSample, Kreckel2013a}, this technique becomes increasingly difficult for more distant galaxies \citep[e.g.][]{Madau1996}. A robust conversion factor between the two dust attenuation measures, such as is calculated in this work, is therefore an invaluable tool for observing distant galaxies.

The paper is structured as follows. Section~\ref{sec:Observations} outlines technical details of the MaNGA survey, the Galaxy Zoo:3D project, and steps taken during data reduction and analysis. Section~\ref{sec:Dust Attenuation Measures} describes the methods implemented to quantify dust attenuation. In Section~\ref{sec:Linking Properties} we present the results and subsequent discussion of this work, and compare the differences between looking globally and locally at the same data. Finally, we propose a new geometry for dust in spiral galaxies in Section~\ref{sec:Dust Geometry} and draw our conclusions in Section~\ref{sec:Conclusions}.

Throughout this paper, we employ a \citet{Chabrier2003GalacticFunction} initial mass function (IMF). Luminosities, masses, and SFRs are calculated assuming a $\rm \Lambda CDM$ cosmology with $\Omega_M = 0.3$, $\Omega_{\Lambda} = 0.7$, and $H_0 = 70 \: \rm km \ s^{-1} \ Mpc^{-1}$.

\section{Observations and Data}
\label{sec:Observations}

\subsection{The MaNGA Survey}
\label{subsec:MaNGA}

The MaNGA galaxy survey \citep{Bundy2015OVERVIEWOBSERVATORY} is part of the fourth generation of the Sloan Digital Sky Survey \citep[SDSS-IV;][]{Blanton2017SloanUniverse}, and will survey over 10~000 nearby galaxies by 2020 \citep{Yan2016SDSS-IVQuality, Wake2017TheConsiderations}. It uses Integral Field Unit (IFU) spectroscopy to obtain detailed spectral information about these galaxies. Hexagonal arrangements of IFU fibres are used to obtain spatial coverage to a distance of 1.5 and 2.5 effective radii ($R_{\rm e}$) for the primary and secondary samples respectively \citep{Law2015OBSERVINGSURVEY}. Each fibre bundle is connected to a multi-object fibre spectrograph \citep{Smee2013THESURVEY, Drory2015THETELESCOPE} attached to the $2.5 \: \rm m$ telescope at Apache Point Observatory \citep[APO;][]{Gunn2006TheSurvey}. MaNGA has a wavelength range of $3600 - 10 \: 300~\si{\angstrom}$, and a spectral resolution of $R \sim 2000$ \citep{Bundy2015OVERVIEWOBSERVATORY}. The result is high-quality optical spectroscopy across the face of a large sample of low-redshift galaxies.

Raw data from MaNGA are calibrated \citep{Yan2016SDSS-IV/MaNGA:TECHNIQUE} and reduced by the Data Reduction Pipeline \citep[DRP;][]{Law2016TheSurvey}; spectral indices and stellar and gas properties of the galaxies, such as the Gaussian profile integrated emission line fluxes, used in this work are made available thanks to the Data Analysis Pipeline \citep[DAP;][]{Westfall2019TheOverview, Belfiore2019TheModeling}. DAP products may be accessed, inspected, and downloaded using the Python package and web application \texttt{Marvin} \citep{Cherinka2018Marvin:Set}. Prior to analysis, the DAP subtracts the stellar spectrum from the emission lines of each MaNGA spaxel, correcting for stellar absorption and providing purely the emission due to the gas. The DAP also accounts for the Milky Way reddening using the \citet{ODonnell1994RSUBnu/SUB-dependentExtinction} reddening law.

The MaNGA sample is inherently mass-biased: high-mass galaxies are over-represented and low-mass galaxies are under-represented in the sample, in order to obtain similar numbers of galaxies in all mass ranges. In certain circumstances, such as when average global quantities for the sample are calculated in this work, it is necessary to correct for this bias to ensure that the sample is representative of the Universe as a whole. This correction is achieved via the use of the DRP \texttt{esweights} to statistically convert the sample into a volume-limited, mass-complete one. We refer to this correction throughout this work as ``volume-weighting''.

Galactocentric radius values used in this work are normalised by the elliptical Petrosian effective radius (a measure of the half-light radius), based on the SDSS $r$-band, from the NASA Sloan Atlas \citep[NSA;][]{Blanton2005NewSurveys, Blanton2011ImprovedImages} catalogue. This method is the most reliable way to measure photometric properties of MaNGA galaxies, and is done in order to account for the effects of inclination on radius measurements. Similarly, stellar mass estimates are taken from the NSA catalogue, again derived from elliptical Petrosian photometry; these masses are produced using \texttt{KCORRECT} \citep{Blanton2006K-correctionsInfrared}, and assuming a \citet{Chabrier2003GalacticFunction} IMF. See \citet{Fraser-McKelvie2019FromGalaxy} for an in-depth discussion on obtaining stellar mass estimates for SDSS galaxies.

\subsection{Sample Selection using Galaxy Zoo:3D}
\label{subsec:Sample Selection}

We analyse only spiral galaxies in this work for several reasons. Firstly, sample selection based purely on SFR or colour could select very different physical systems. However, studying only disk galaxies and excluding irregular galaxies from our analysis means that the distribution of stars, gas and dust within the galaxies is more regular, with a clearly defined disk and spiral arms. Furthermore, we also wanted to test whether dust attenuation properties between the spiral arm and inter-arm regions of the galaxies exhibit any differences.

Since we are endeavouring to analyse a well-defined sample of unequivocally spiral galaxies, we choose a subset of 253 galaxies classified as spirals by Galaxy Zoo:3D (GZ:3D; Masters et al., \emph{in prep.}), which are drawn from the primary and secondary samples of the eighth MaNGA Product Launch (MPL-8) data release.\footnote{Although we here use data from MPL-8, the GZ:3D sample was originally a subset of the MPL-5 spirals which more than 50\% of respondents to the Galaxy Zoo 2 \citep[GZ2;][]{Willett2013GalaxySurvey, Hart2016GalaxyBias} project marked as having either spiral arms or a bar.} GZ:3D is a citizen science project in which volunteers identify the spiral arms and bars of MaNGA galaxies using a freehand drawing tool, which people use to decide how best to enclose the spiral arms of galaxies in SDSS images. Each person effectively `votes' on the MaNGA spaxels which they consider to be part of the spiral arms. The result is a mask for the 253 spiral galaxies: spaxels with a greater number of votes are weighted accordingly and are likely to comprise part of the spiral arms of the galaxies.

Selecting only the GZ:3D spirals has two further advantages. First, almost all GZ:3D spirals are close to face-on, as a result of volunteers determining the location of spiral arms more readily in face-on galaxies than in edge-on galaxies. Consequently, it is easier to consistently compare dust attenuation properties of different galaxies within this sample, since the variation of inclination from galaxy to galaxy is small. Moreover, choosing purely GZ:3D spirals ensures that each galaxy has well-defined spiral arms, reducing ambiguity in subsequent analysis.

We excluded any galaxies with DAP datacube quality flags cautioning ``do not use'', or with two or more ``warning'' flags, which reduced the sample to 247 galaxies. Finally, to ensure that the individual spaxels used for analysis in this work were of a sufficiently high quality, a minimum continuum signal to noise (S/N) cut of 10 or higher was imposed; spaxels below this threshold were discarded prior to analysis.

\subsection{BPT Diagrams}
\label{subsec:BPT diagrams}

The Balmer decrement is only a valid measure of the dust attenuation due to the clumpy BCs associated with the youngest stellar populations if the $\rm H \alpha$ and $\rm H \beta$ emission lines for a given spaxel are actually produced by the recombination of gas ionised by these stars. This is because the adopted intrinsic value of the ratio of the $\rm H \alpha$ emission line to the $\rm H \beta$ line is dependent on the emission being produced by photoionisation. We therefore analyse only regions of the galaxies that are dominated by SF processes, and exclude regions for which active galactic nuclei (AGN) -- and other excitation mechanisms -- are responsible for the majority of the line emission. To this end, we employ the most commonly used Baldwin, Phillips and Terlevich (BPT; \citealp{Baldwin1981ClassificationObjects}; see also \citealp{Veilleux1987SpectralGalaxies}) diagram. This diagnostic compares the line ratio ${\rm [\textsc{O~iii}] \: \lambda 5007 / H\beta}$ to ${\rm [\textsc{N~ii}] \: \lambda 6583 / H\alpha}$. We therefore excluded galaxies which only had $\rm H\alpha$, $\rm H\beta$, ${\rm [\textsc{O~iii}] \: \lambda 5007}$ or ${\rm [\textsc{N~ii}] \: \lambda 6583}$ signal data for $< 5\%$ of their spaxels, which reduced the sample yet further to 232 galaxies. Furthermore, individual spaxels lacking signal data for any of the above emission lines were also rejected.

This analysis follows the work of \citet{Belfiore2015P-MaNGAObservations, Belfiore2016SDSSLIERs}, who also provide an overview of BPT diagrams in their work. The line ratios described above are particularly useful for this work because they have similar wavelengths, and so will not be affected much by the dust that we are endeavouring to measure. The exact demarcation between SF and non-SF regions is not uniquely defined; in this work, we follow the method outlined by \citet{Belfiore2016SDSSLIERs} for applying BPT diagrams to MaNGA galaxies and consider line emission in spaxels below the less stringent \citet{Kewley2001TheoreticalGalaxies} demarcation line to be dominated by SF regions. Consequently, spaxels in the `composite' region between the demarcation lines of \citet{Kewley2001TheoreticalGalaxies} and \citet{Kauffmann2003TheNuclei} are classified as SF in this work. Spaxels above the \citet{Kauffmann2003TheNuclei} line are assumed to be dominated by excitation mechanisms that are not SF in nature -- such as AGN, Low Ionisation (Nuclear) Emission line Regions (LI(N)ERs), shocks and Seyfert galaxies -- and are thus discarded prior to analysis. In total, the final sample used in this work comprises almost $140 \: 000$ spaxels, belonging to 232 SF spiral galaxies out of the original 253 GZ:3D galaxies.

\section{Measures of Dust Attenuation}
\label{sec:Dust Attenuation Measures}

\subsection{Measuring Dust Attenuation from Emission Line Ratios}
\label{subsec:Balmer Decrement}

\begin{figure*}
	\includegraphics[width=\textwidth]{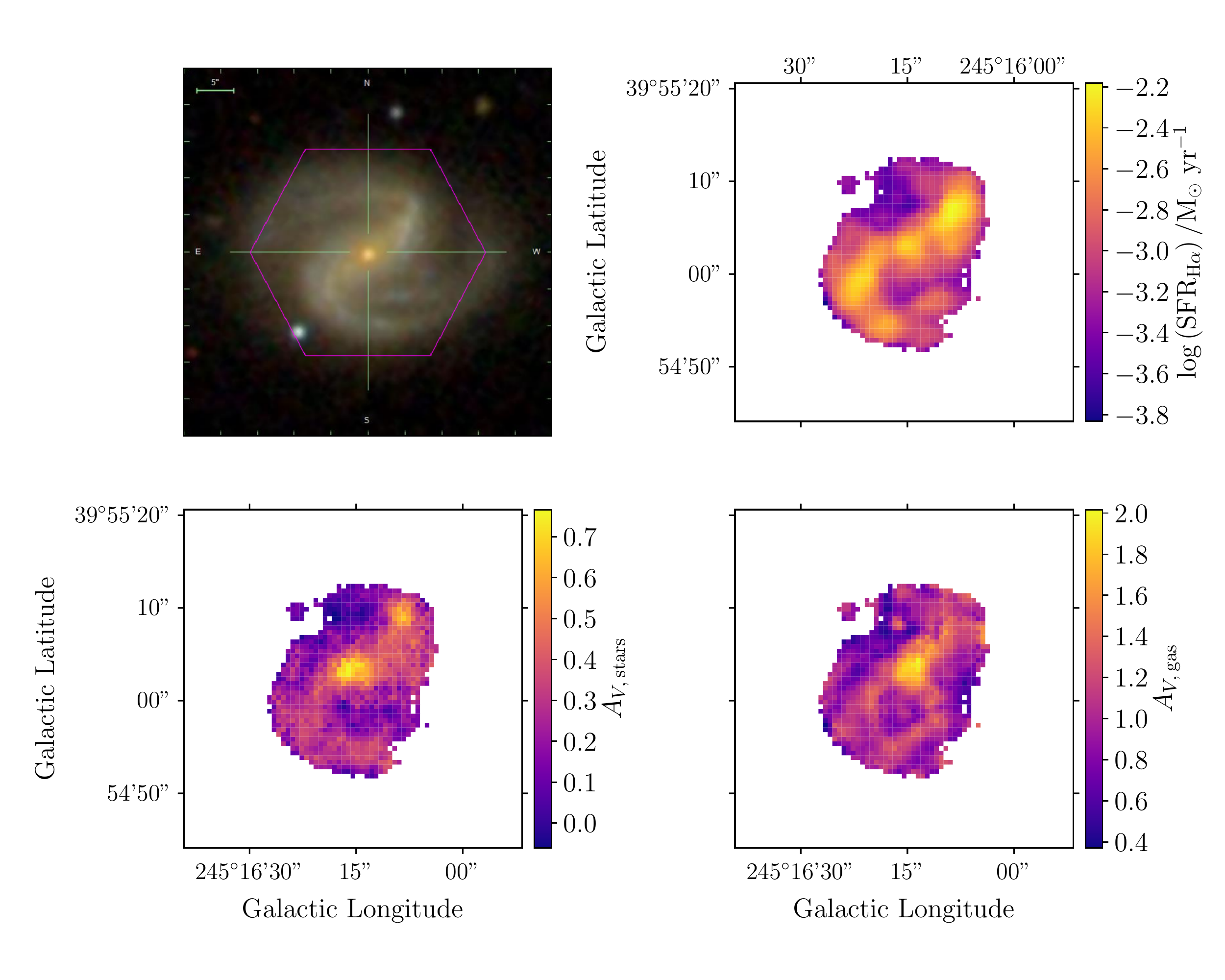}
    \caption{3-colour SDSS image (top left), Balmer-decrement corrected SFR map (top right), dust attenuation map produced by \texttt{STARLIGHT} (bottom left), and dust attenuation map derived from the Balmer decrement (bottom right) for galaxy 8312-12702.}
    \label{fig:spiral_example}
\end{figure*}

The amount of dust attenuation along a given line of sight may be inferred by measuring the ratio of the $\rm H \alpha$ emission line to the $\rm H \beta$ line of the spectrum concerned. This ratio, known as the Balmer decrement, is fundamentally determined both by quantum mechanics and the physical conditions of the gas (namely its electron temperature and electron density). Any measured deviation from its intrinsic value can therefore be attributed to dust attenuation for a given electron temperature and electron density, assuming Case B recombination \citep{Osterbrock2005AstrophysicsNuclei}. See \citet{Groves2012a} for a comprehensive discussion on the Balmer decrement pertaining to SDSS galaxies.

Ultimately, we want to produce Balmer-decrement derived dust attenuation maps for the sample of galaxies, as well as Balmer-decrement corrected SFR maps, so that we can categorise spaxels by SFR per unit physical surface area. As described in Section~\ref{subsec:MaNGA}, the initial processing of MaNGA data by the DAP has already corrected the emission line fluxes for Milky Way reddening and absorption in the underlying stellar spectrum. Therefore, we need only correct for the dust attenuation intrinsic to the galaxies we are analysing. We first calculate the colour excess $E(B-V)$ using the equation from \citet{Osterbrock2005AstrophysicsNuclei}, as in \citet{Dominguez2013DUSTSURVEY, Kreckel2013a}:

\begin{equation}
    E(B-V) = 1.97 \log_{10} \left[ \frac{{\rm \left( H \alpha / H \beta \right)_{obs}}}{2.87} \right].
	\label{eq:colour_excess}
\end{equation}
Equation~(\ref{eq:colour_excess}) assumes an intrinsic value for ${\rm \left( H \alpha / H \beta \right)_{int}} = 2.87$, which is valid for an electron temperature, $T_{\rm e}$, of $T_{\rm e} = 10^4 \: \rm K$ and an electron density, $n_{\rm e}$, of $n_{\rm e} = 10^2 \: \rm cm^{-3}$, under Case B recombination conditions \citep{Osterbrock2005AstrophysicsNuclei}.

${\rm \left( H \alpha / H \beta \right)_{int}}$ is not constant within galaxies; it is dependent upon both the local electron temperature and density. However, as discussed in Section~\ref{subsec:BPT diagrams}, we only analyse those regions of the galaxies for which SF is the dominant source of line emission. In these regions, \citet{Osterbrock2005AstrophysicsNuclei} find that at constant $n_{\rm e} = 10^2 \: \rm cm^{-3}$, ${\rm \left( H \alpha / H \beta \right)_{int}}$ varies between 3.05 for an electron temperature of $T_{\rm e} = 5 \times 10^3 \: \rm K$, and 2.76 for $T_{\rm e} = 2 \times 10^4 \: \rm K$. The dependence on electron density is weaker: at constant $T_{\rm e} = 10^4 \: \rm K$, ${\rm \left( H \alpha / H \beta \right)_{int}}$ falls only to 2.85 at $n_{\rm e} = 10^4 \: \rm cm^{-3}$, and to 2.81 at $n_{\rm e} = 10^6 \: \rm cm^{-3}$ \citep{Osterbrock2005AstrophysicsNuclei}. Altering the value of ${\rm \left( H \alpha / H \beta \right)_{int}}$ between these extreme values of 3.05 and 2.76 results in an uncertainty in $E(B-V)$ of less than 0.1. Since such a variation is negligible for the analysis in this work, we therefore assume that each spaxel has a uniform value of ${\rm \left( H \alpha / H \beta \right)_{int}} = 2.87$.

The ratio of the dust attenuation at some wavelength $\lambda$, $A_{\lambda}$, to the colour excess is denoted $R_{\lambda} = A_{\lambda} / E(B-V)$, and is determined at that wavelength for a given reddening curve. In this work, we adopt the reddening curve of \citet{Calzetti2000TheGalaxies}. When deriving the dust attenuation in the gas from measurements of the Balmer decrement, we choose $R_V = 3.1$; when measuring the dust attenuation affecting the stellar populations, however, we choose $R_V = 4.05$. For a detailed explanation as to why different $R_V$ values are used to determine the two attenuation measures, we refer the reader to \citet{Calzetti2000TheGalaxies}; see also \citet{Catalan-Torrecilla2015StarData, Pannella2015GOODS-HERSCHEL:4}. For the gas, we thus obtain $A_{V, \: \rm gas} = 3.1 E(B-V)$, and for the stellar populations, we instead have $A_{V, \: \rm stars} = 4.05 E(B-V)$. In this way, the dust attenuation can be calculated purely from measurements of the emission line fluxes of $\rm H\alpha$ and $\rm H\beta$ for each SF spaxel within all galaxies in the sample.

\subsection{Measuring Dust Attenuation from Stellar Population Modelling}
\label{subsec:Stellar Populations}

We also determine an entirely independent measure of the dust attenuation affecting the stellar populations at each location on the galaxy. We use the full-spectrum stellar population fitting code \texttt{STARLIGHT} \citep{CidFernandes2005Semi-empiricalMethod} to derive a best-fit spectrum, created from a linear combination of a set of input single stellar population (SSP) template spectra -- with no assumptions or restrictions on the shape of the derived star formation histories -- and an applied dust attenuation. The fitting method used is similar to that used by \citet{Peterken2019Time-slicingMaNGA}, and is described fully by Peterken et al. (\emph{in prep.}). Here we highlight the key relevant details.

Before fitting each spaxel's spectrum, emission lines are removed using the MaNGA DAP \citep{Westfall2019TheOverview, Belfiore2019TheModeling}. No binning of neighbouring spaxels is done, since we wish to fully retain all spatial information. \citet{Ge2018RecoveringUncertainties} show that \texttt{STARLIGHT} can be susceptible to significant biases at low S/N, so we use the \texttt{STARLIGHT} configuration settings subsequently recommended by \citet{CidFernandes2018OnAlgorithms}, which prioritise robustness against such biases over computation times. 54 of the SSPs used are the E-MILES (Extended Medium-resolution Isaac Newton Telescope Library of Empirical Spectra) templates of \citet{Vazdekis2016UV-extendedGalaxies}, based on the earlier MILES library \citep{Vazdekis2010EvolutionarySystem}. These templates cover a range of nine ages (in the range $7.85 \leq \log(\rm age / yr) \leq 10.25$) and six metallicities ($-1.71 \leq [\rm M / H] \leq +0.22$); they assume a \citet{Chabrier2003GalacticFunction} IMF, ``Padova'' isochrones \citep{Girardi1999Evolutionary0.03}, and a Milky Way [$\alpha / \rm Fe$] (``baseFe''). As well as these E-MILES templates, we also include a further six ages (in the range $6.8 \leq \log(\rm age / yr) \leq 7.6$) and two metallicities ($[\rm M / H] = -0.41, +0.00$) from the \citet{Asad2017YoungCMDs} extension to the E-MILES library. These younger templates assume \citet{Bertelli1994TheoreticalOpacities.} isochrones, and are very similar to those of \citet{Girardi1999Evolutionary0.03} since they are also part of the Padova project; they are generated using the same method as the E-MILES SSP templates.

When building a best-fit spectrum from the SSPs, \texttt{STARLIGHT} also fits a dust attenuation $A_{V, \: \rm stars}$, for which a \citet{Calzetti2000TheGalaxies} attenuation curve is assumed with an $R_V$ of $4.05$ \citep{Calzetti2000TheGalaxies, Catalan-Torrecilla2015StarData, Pannella2015GOODS-HERSCHEL:4}, as discussed in Section~\ref{subsec:Balmer Decrement}. The large wavelength range of the MaNGA and E-MILES spectra are exploited by fitting between 3541.4 and 8950.4~\AA\ to use as much of the information contained in the whole continuum shape as possible to constrain the value of $A_{V, \: \rm stars}$. The value of $A_{V, \: \rm stars}$ is allowed to vary in the range $-1 \leq A_{V, \: \rm stars} \leq 8$ to allow for a full exploration of reasonable parameter space. The best fit results are, reassuringly, nearly all contained within the range $0 \leq A_{V, \: \rm stars} \leq 0.8$.

The \texttt{STARLIGHT} fit only applies a single $A_{V, \: \rm stars}$ to the overall spectrum. Such an approach is a simplification, since we expect the spectra of the youngest stellar populations to be attenuated to a greater degree than those of older populations. To allow for this discrepancy, \texttt{STARLIGHT} includes an option to include extra attenuation to be applied to specified SSP templates in the fit. However, in practice, the youngest SSPs do not have strong spectral features, making this extra attenuation degenerate with changes in the continuum shape with population age or metallicity. We therefore do not use this feature. In addition, the contribution of the youngest stellar populations to the total spectrum is difficult to distinguish from the presence of hot stars present -- but not modelled -- in the oldest SSP templates which are responsible for the ``UV~upturn'' (see \citealp{Yi2008TheUpturn} for a review). Peterken et al. (\emph{in prep.}) show that this results in \texttt{STARLIGHT} fits which are unreliable for stellar populations younger than 30~Myr (an effect also found by \citealp{CidFernandes2009TestingMethodology}), but that the older populations are still modelled well.

Together, these limitations mean that the \texttt{STARLIGHT} attenuation measurements are liable to some uncertainties regarding the attenuation of the youngest stars. However, the value $A_{V, \: \rm stars}$ used in the fit is likely to be a good indicator of the average attenuation of the flux from the average stellar population. The stellar population effectively probed by \texttt{STARLIGHT}'s attenuation measurement is significantly older than that probed using the Balmer decrement, which specifically measures the attenuation in the star-forming regions.

These $A_{V, \: \rm stars}$ measurements are determined for every SF spaxel in each of the sample galaxies. An example comparing the dust attenuation obtained in this way to that derived from measurements of the Balmer decrement (described in Section~\ref{subsec:Balmer Decrement}) for galaxy 8312-12702 is shown in Fig.~\ref{fig:spiral_example}.

\section{Linking Dust and Star Formation Properties}
\label{sec:Linking Properties}

In order to understand whether variations in dust attenuation are driven primarily by global or local properties of galaxies, the data in this work are analysed on three different scales. Firstly, we categorise entire galaxies based on global quantities. We next take a more detailed look at the local quantities within these global categories by making use of the spatial resolution of MaNGA. Finally, we look yet more locally at the data, categorising spaxels rather than galaxies into different categories depending on local criteria.

\subsection{Categorising Galaxies by Global Properties}
\label{subsec:Categorising galaxies}

\begin{figure}
	\includegraphics[width=0.5\textwidth]{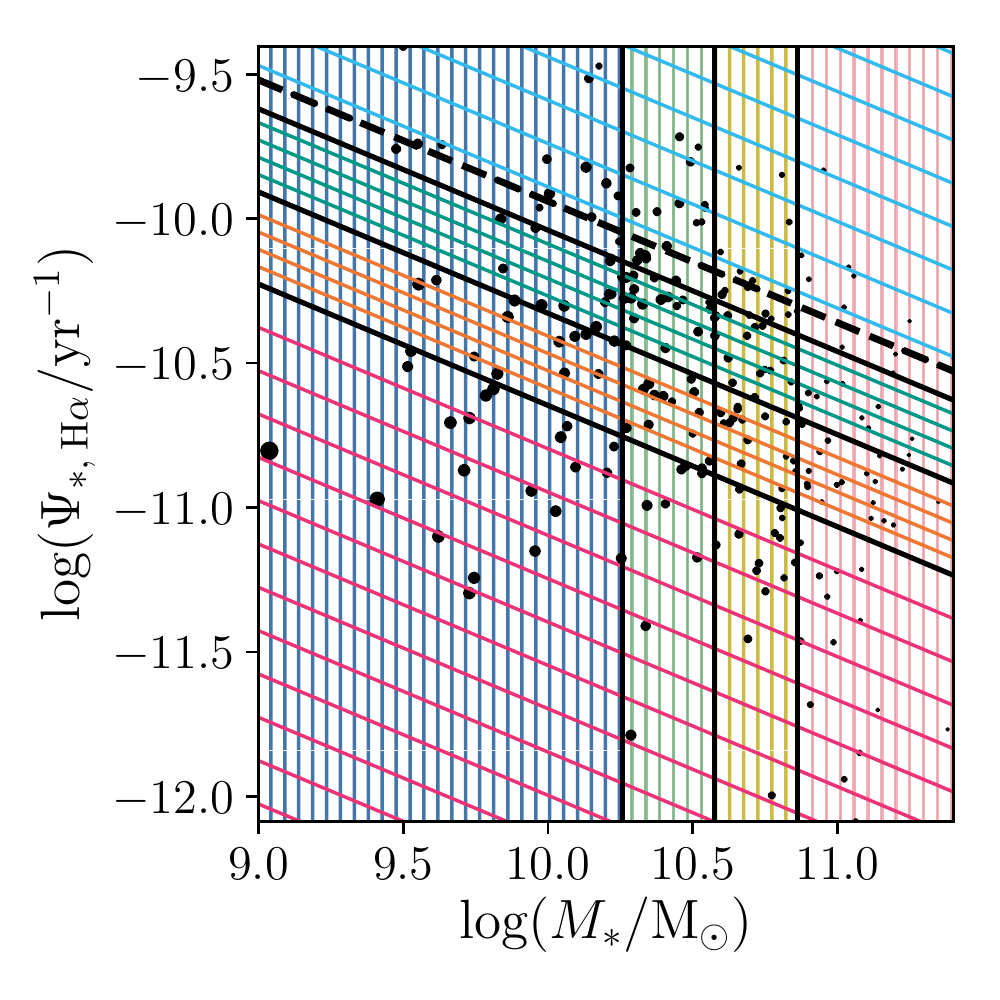}
    \caption{Specific SFR relative to the SF main sequence (dashed black line; \citealp{Grootes2017GalaxyGalaxies}) versus stellar mass for galaxies in the sample. Galaxies are first categorised based on their mass; we choose there to be roughly the same number of galaxies within each category. Galaxies in the blue region have stellar masses of ${\rm log} \left(M_{*} / \rm{M_{\odot}} \right) < 10.26$, those in the green have $10.26 < {\rm log} \left(M_{*} / \rm{M_{\odot}} \right) < 10.58$, those in the yellow have $10.58 < {\rm log} \left(M_{*} / \rm{M_{\odot}} \right) < 10.86$, and those in the red have ${\rm log} \left(M_{*} / \rm{M_{\odot}} \right) > 10.86$. We subsequently categorise galaxies by their vertical offset from the SFMS, $\Delta {\rm log} \left(\Psi_{*, \: \rm H \alpha} / \rm{yr^{-1}} \right)$: galaxies in the magenta region have a $\rm sSFR_{H \alpha}$ of $\Delta {\rm log} \left(\Psi_{*, \: \rm H \alpha} / \rm{yr^{-1}} \right) < -0.71$ from the SFMS, those in the orange have $-0.71 < \Delta {\rm log} \left(\Psi_{*, \: \rm H \alpha} / \rm{yr^{-1}} \right) < -0.39$ from the SFMS, those in the teal have $-0.39 < \Delta {\rm log} \left(\Psi_{*, \: \rm H \alpha} / \rm{yr^{-1}} \right) < -0.10$ from the SFMS, and those in the cyan have $\Delta {\rm log} \left(\Psi_{*, \: \rm H \alpha} / \rm{yr^{-1}} \right) > -0.10$ from the SFMS. Solid black lines show the boundaries between these regions. The size of each point is directly proportional to the volume weight for each galaxy. Note that this plot does not show the total sSFRs of the galaxies: simply the sum of $\rm SFR_{H \alpha}$ across all SF spaxels within each of the galaxies divided by the total stellar mass.}
    \label{fig:galaxy_types}
\end{figure}

We first present and discuss the results obtained from looking just at the global properties of the galaxies. In their seminal paper, \citet{Calzetti2000TheGalaxies} found empirically that the global ratio of the reddening -- i.e. the colour excess $E(B-V)$ -- affecting the stellar continuum to that derived from the Balmer decrement was equal to $0.44 \pm 0.03$. Since the observed flux from the emission lines and the stellar continuum is not produced by the same stars, the attenuation of the former is more than twice as high as that of the latter \citep{Keel1993ObscurationNuclei, Calzetti1994DustLaw}. The data of \citet{Calzetti2000TheGalaxies} were obtained by the International Ultraviolet Explorer \citep[IUE;][]{Boggess1978TheInstrumentation}. The IUE possessed both a high- and low-resolution spectrograph but was not equipped for integral field spectroscopy \citep{Boggess1978TheInstrumentation}. Therefore, we initially consider just the global properties of the galaxies in the sample; we do not (yet) make use of the spatial information that MaNGA provides for each galaxy.

Although our initial objective was simply to measure the value of this ratio obtained by these authors, we also ultimately want to determine which galactic properties are the main drivers of variations in dust attenuation. Therefore, we choose to categorise the sample into four types by looking at two important global properties of galaxies (as shown in Fig.~\ref{fig:galaxy_types}) -- first their stellar mass, and second their specific SFR (sSFR; defined as SFR / stellar mass) relative to the star-forming main sequence (SFMS; \citealp{Grootes2017GalaxyGalaxies}). Of course, there are myriad SFMS calibrations from which to choose (for instance \citealp{Noeske2007StarGalaxies, Peng2010MASSFUNCTION, Whitaker2012THE2.5, Guo2015THEMASS, RodriguezdelPino2017OMEGAActivity} -- and references within, notably \citealp{Brinchmann2004TheUniverse} and \citealp{Abazajian2009THESURVEY} -- \citealp{Spindler2017SDSS-IVEnvironment, Grootes2017GalaxyGalaxies, Salim2018DustAnalogs}). We employ the SFMS calculated by \citet{Grootes2017GalaxyGalaxies} since these authors fit a SFMS for spiral galaxies with redshifts of $z < 0.06$, which is the appropriate redshift range for the vast majority of the galaxies analysed in this work. \citet{Grootes2017GalaxyGalaxies} derive SFRs by integrating the total near-UV luminosity of the galaxies in their sample.

In order to produce SFR maps for each of the galaxies, the intrinsic $\rm H \alpha$ flux $f_{\rm int, \: H \alpha}$ is determined for every SF spaxel in each of the galaxies in the sample:

\begin{equation}
    f_{\rm int, \: H \alpha} = f_{\rm obs, \: H \alpha} \times 10^{0.4 A_{\rm H \alpha}},
	\label{eq:intrinsic_flux}
\end{equation}
where $f_{\rm obs, \: H \alpha}$ is the raw $\rm H \alpha$ flux from the DAP, and the $\rm H \alpha$ dust attenuation, $A_{{\rm H \alpha}} = 2.38 E(B-V)$ for SF galaxies (using the \citealp{Calzetti2000TheGalaxies} reddening curve with $R_V = 3.1$). The $\rm H \alpha$ luminosity in Watts, $L_{{\rm H \alpha}} \: {\rm \left[ W \right]}$, of the galaxies is determined using the luminosity distances provided by the NSA catalogue \citep{Blanton2005NewSurveys, Blanton2011ImprovedImages} for our chosen cosmology. The $\rm H \alpha$ derived SFR, $\rm SFR_{H \alpha}$, for each SF spaxel in the galaxies may then be calculated using the \citet{Kennicutt1998TheGalaxies} relation, assuming a \citet{Chabrier2003GalacticFunction} IMF -- see also \citet{Kreckel2013a, Schaefer2017TheGalaxies, Schaefer2018TheGroups}:

\begin{equation}
    {\rm SFR_{H \alpha}} \: \left[{\rm M_\odot \ yr^{-1}} \right] = \frac{L_{{\rm H \alpha}} \: {\rm \left[ W \right]}}{2.16 \times 10^{34} \: \rm M_\odot \ yr^{-1}}.
	\label{eq:SFR}
\end{equation}
A $\rm SFR_{H \alpha}$ map produced in this way for one of the galaxies in the sample is shown in Fig.~\ref{fig:spiral_example}.

We first split the galaxies into four distinct categories based on their stellar masses (obtained from the NSA catalogue; \citealp{Blanton2005NewSurveys, Blanton2011ImprovedImages}), as seen in Fig.~\ref{fig:galaxy_types}. These masses are also used in conjunction with the $\rm H \alpha$ derived SFRs calculated from Equation~(\ref{eq:SFR}) to determine the global $\rm sSFR_{H \alpha}$, $\Psi_{*, \: \rm H \alpha}$, for each galaxy. The galaxies are then additionally categorised by their vertical offset from the SFMS, $\Delta {\rm log} \left(\Psi_{*, \: \rm H \alpha} / \rm{yr^{-1}} \right)$ (shown in Fig.~\ref{fig:galaxy_types}; \citealp{Grootes2017GalaxyGalaxies}); that is to say, galaxies are categorised by their $\rm sSFR_{H \alpha}$ at a given mass. We choose not to categorise galaxies by $\rm sSFR_{H \alpha}$ alone, since doing so would introduce bias due to the implicit dependence on stellar mass.

Figure~\ref{fig:galaxy_types} does not actually show the total sSFRs of the galaxies, since Balmer-decrement corrected SFRs can be calculated only for SF spaxels, as discussed in Section~\ref{subsec:BPT diagrams}. Moreover, and of greater concern, the spatial coverage of MaNGA is limited to a distance of $1.5 \ R_{\rm e}$ and $2.5 \ R_{\rm e}$ for the galaxies in the primary and secondary samples, respectively. This means that we are at risk of calculating lower limits on the global $\rm H \alpha$ derived sSFRs of the sample galaxies. Nevertheless, the majority of star formation should still be captured, since it is likely to be enclosed within $1.5 \ R_{\rm e}$.

To estimate how much flux we typically miss in the outer regions of the galaxies, we compared global $\rm SFR_{H \alpha}$ calculated from the primary and the secondary sample galaxies separately. The mean global $\rm SFR_{H \alpha}$ for the primary sample galaxies is $1.5 \pm 0.1 \: \rm M_{\odot} \ yr^{-1}$, whereas the increased spatial coverage means that the mean global $\rm SFR_{H \alpha}$ for the secondary sample galaxies is slightly higher, at $1.8 \pm 0.3 \: \rm M_{\odot} \ yr^{-1}$. It therefore seems likely that the limited MaNGA field of view means that the global $\rm H \alpha$ derived SFRs calculated in this work are systematically underestimated by ${\sim} 20 \%$. This can be seen in Fig.~\ref{fig:galaxy_types}, in which many of the SF galaxies lie below the SFMS. It should be noted that this systematic effect does not compromise the final results of this work in any way, because we are interested only in analysing relative differences in star formation activity.

Figure~\ref{fig:galaxy_types} therefore plots the sum of the $\rm SFR_{H \alpha}$ of each SF spaxel for each galaxy divided by the total stellar mass versus the total stellar mass of that galaxy. Bins are chosen such that there are roughly the same number of galaxies within each category. The lowest mass galaxies (blue) have stellar masses of ${\rm log} \left(M_{*} / \rm{M_{\odot}} \right) < 10.26$; the highest mass galaxies (red) have ${\rm log} \left(M_{*} / \rm{M_{\odot}} \right) > 10.86$. Galaxies with the lowest global $\rm H \alpha$ derived sSFRs relative to the SFMS (magenta) have a $\rm sSFR_{H \alpha}$ of $\Delta {\rm log} \left(\Psi_{*, \: \rm H \alpha} / \rm{yr^{-1}} \right) < -0.71$ from the SFMS; those with the highest relative $\rm H \alpha$ derived sSFRs (cyan) have a $\rm sSFR_{H \alpha}$ of $\Delta {\rm log} \left(\Psi_{*, \: \rm H \alpha} / \rm{yr^{-1}} \right) > -0.10$ from the SFMS. The definitions of these categories as outlined above are used when discussing the results of both Section~\ref{subsec:Investigating Global Effects} and Section~\ref{subsec:Looking Locally}.

\subsection{Investigating the Impact of Global Effects on Dust Attenuation}
\label{subsec:Investigating Global Effects}

\begin{figure*}
\centering
	\includegraphics[width=0.3\textwidth]{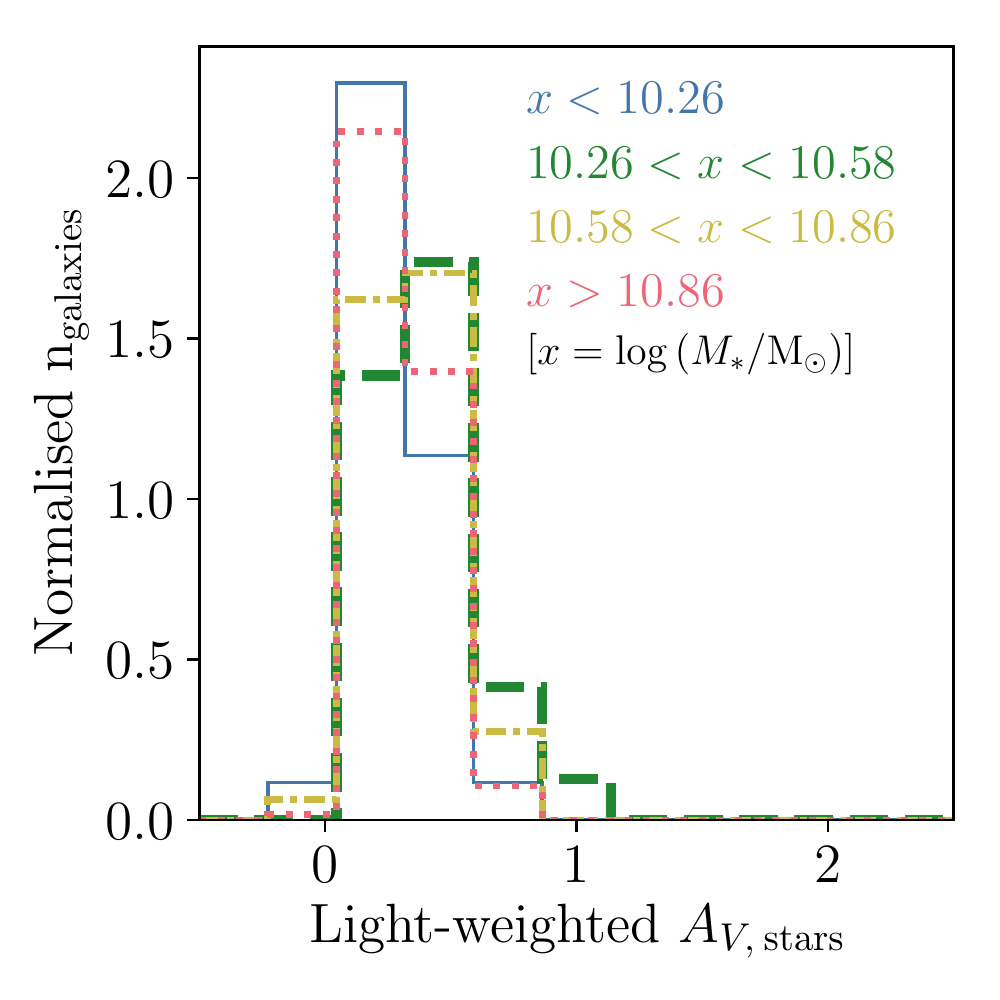}
	\includegraphics[width=0.3\textwidth]{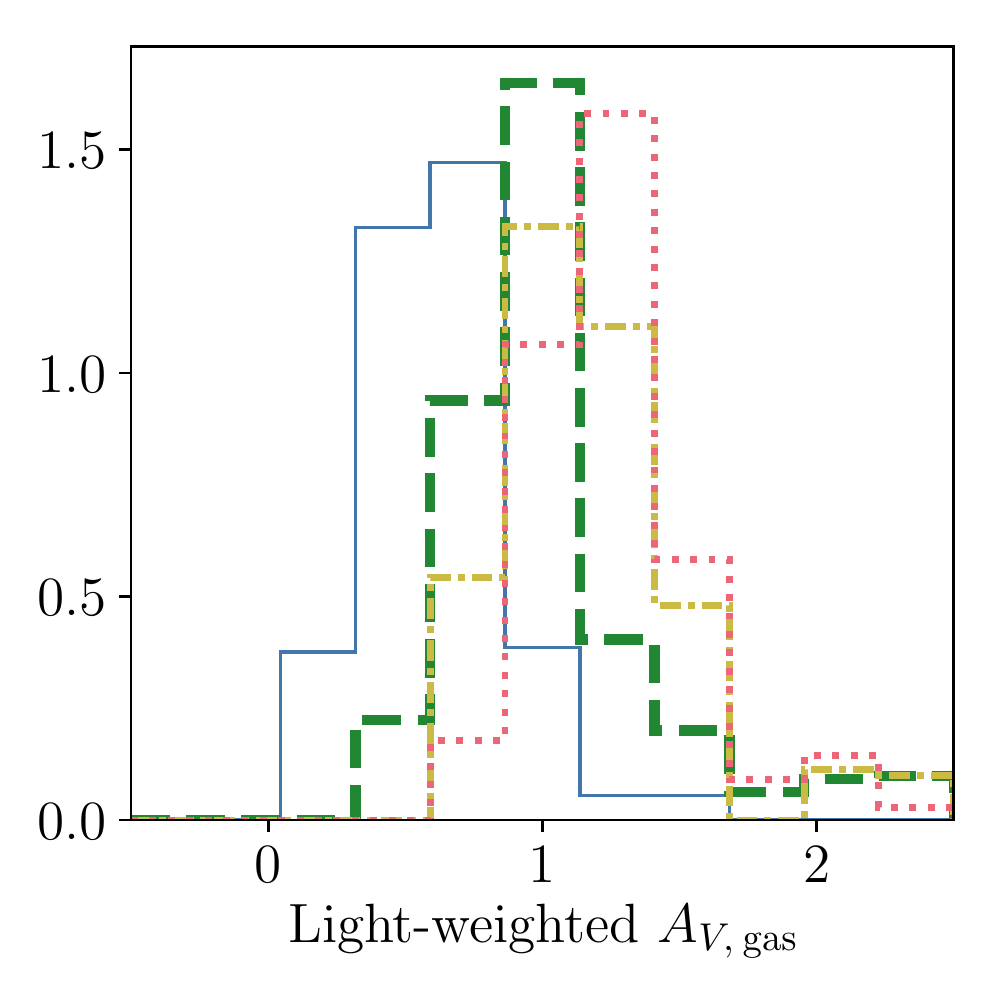}
	\includegraphics[width=0.3\textwidth]{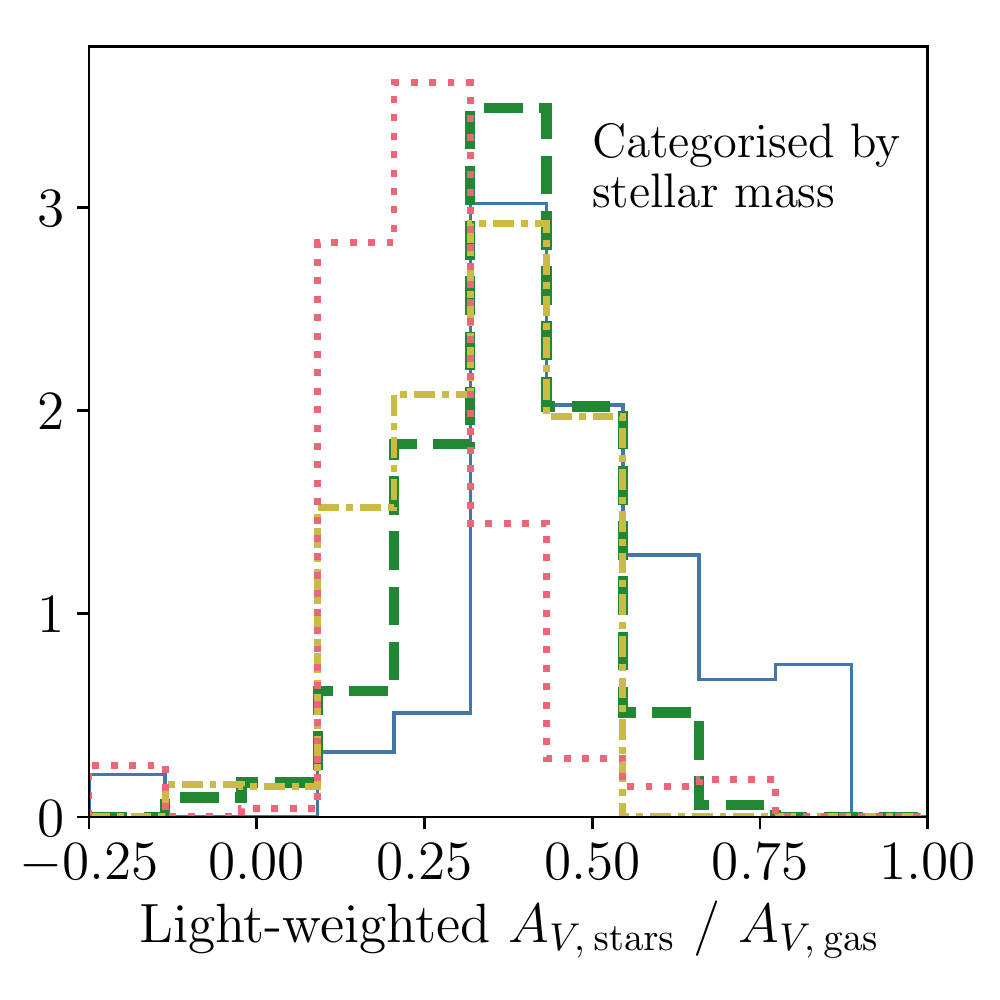}
	\includegraphics[width=0.3\textwidth]{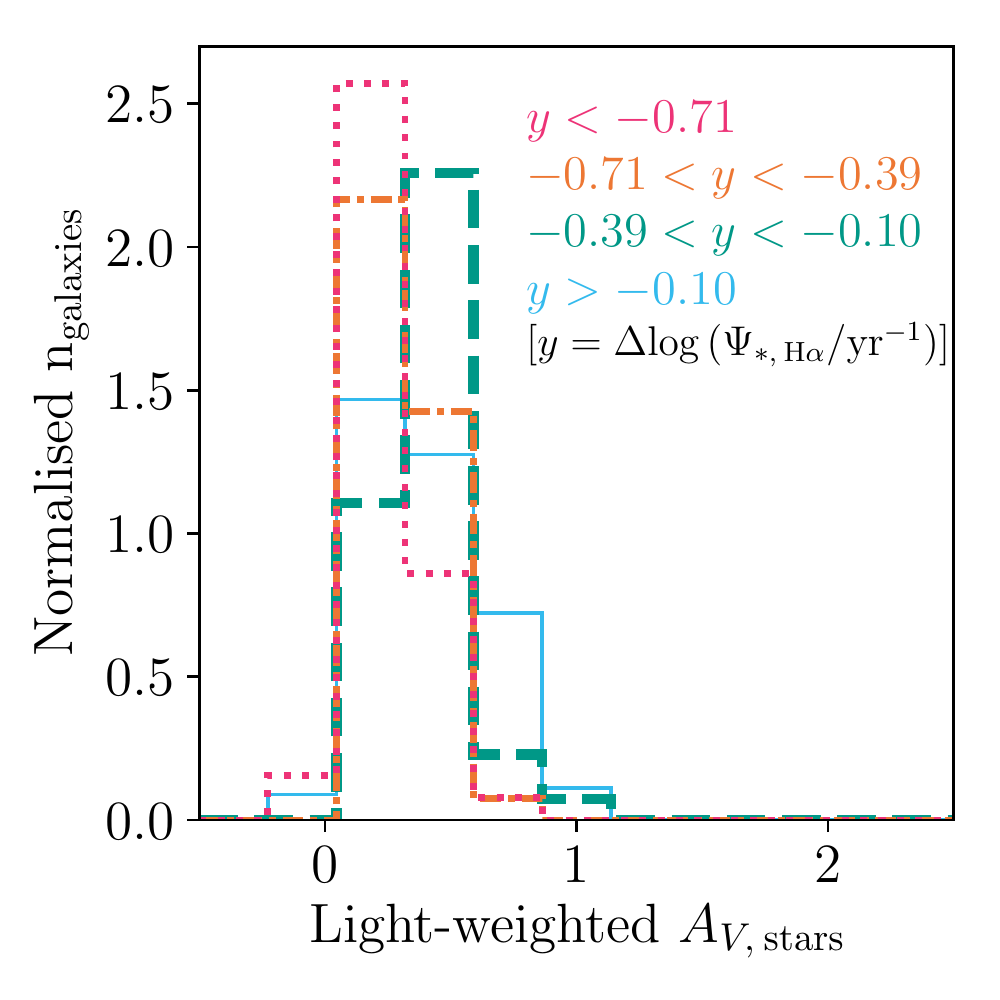}
	\includegraphics[width=0.3\textwidth]{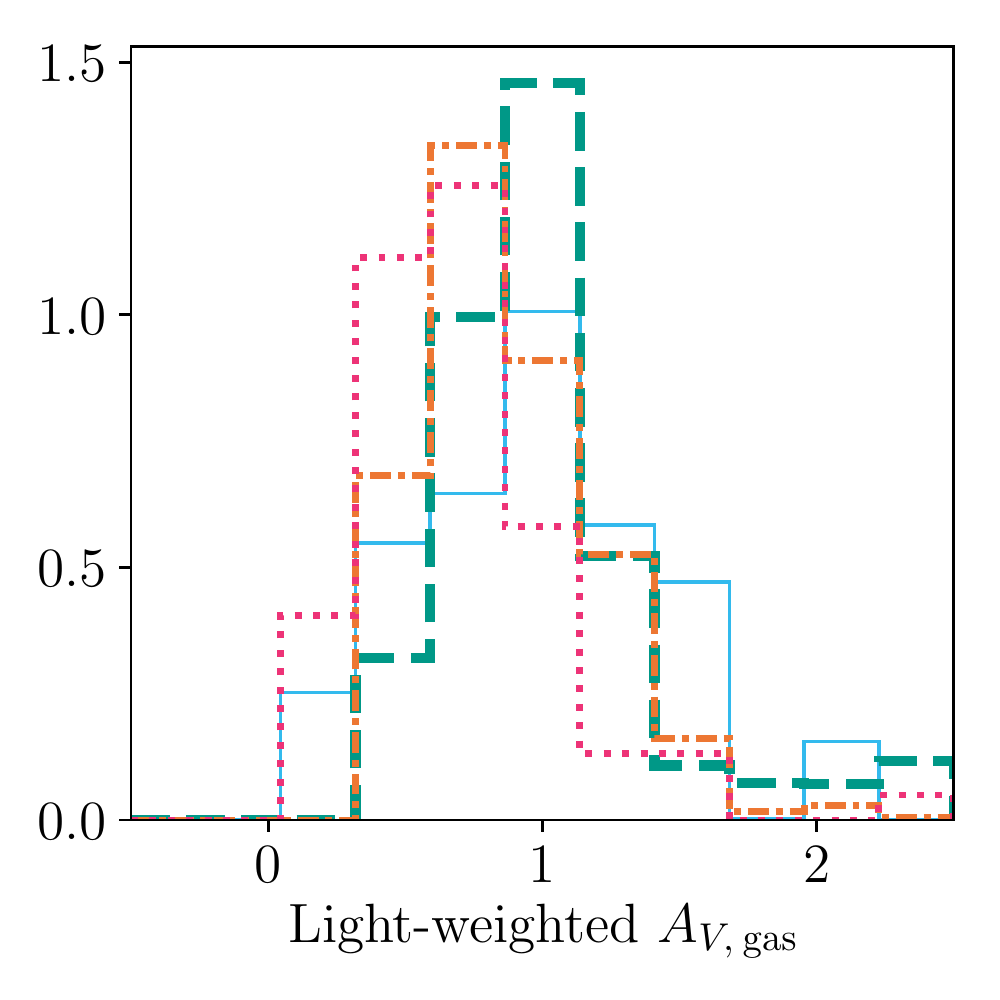}
	\includegraphics[width=0.3\textwidth]{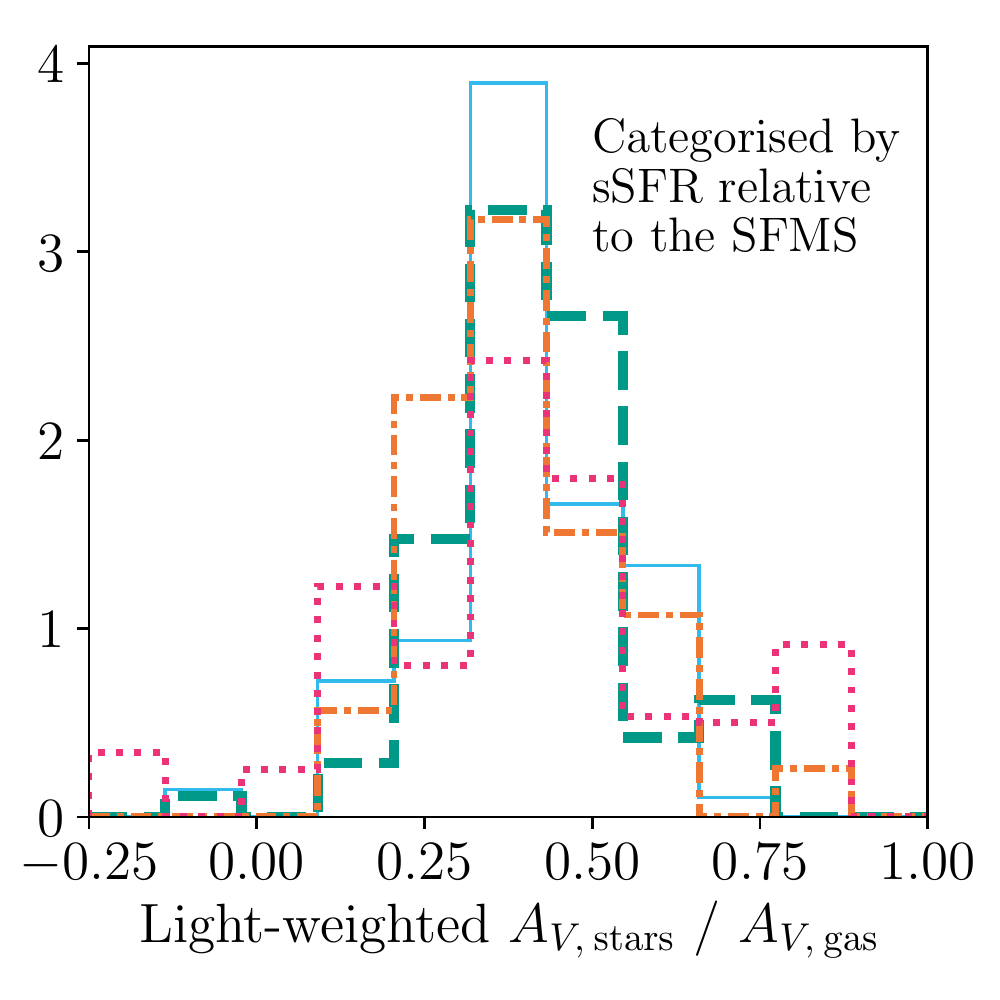}
    \caption{Distributions of (from left to right) light-weighted $A_V$ dust attenuation affecting the stellar populations, light-weighted $A_V$ dust attenuation derived from the Balmer decrement, and the ratio of the two quantities. The bins have been volume-weighted, and are normalised so that the area under each line is equal to one. Distributions on the top row have been categorised by stellar mass; blue lines denote the lowest-mass galaxies with stellar masses of ${\rm log} \left(M_{*} / \rm{M_{\odot}} \right) < 10.26$, green lines denote stellar masses of $10.26 < {\rm log} \left(M_{*} / \rm{M_{\odot}} \right) < 10.58$, yellow lines of $10.58 < {\rm log} \left(M_{*} / \rm{M_{\odot}} \right) < 10.86$, and red lines denote the highest-mass galaxies with stellar masses of ${\rm log} \left(M_{*} / \rm{M_{\odot}} \right) > 10.86$. Distributions on the bottom row have been categorised by global $\rm sSFR_{H \alpha}$ relative to the SFMS; magenta lines denote that galaxies have a $\rm sSFR_{H \alpha}$ of $\Delta {\rm log} \left(\Psi_{*, \: \rm H \alpha} / \rm{yr^{-1}} \right) < -0.71$ from the SFMS, orange lines denote a $\rm sSFR_{H \alpha}$ of $-0.71 < \Delta {\rm log} \left(\Psi_{*, \: \rm H \alpha} / \rm{yr^{-1}} \right) < -0.39$ from the SFMS, teal lines denote a $\rm sSFR_{H \alpha}$ of $-0.39 < \Delta {\rm log} \left(\Psi_{*, \: \rm H \alpha} / \rm{yr^{-1}} \right) < -0.10$ from the SFMS, and cyan lines denote a $\rm sSFR_{H \alpha}$ of $\Delta {\rm log} \left(\Psi_{*, \: \rm H \alpha} / \rm{yr^{-1}} \right) > -0.10$ from the SFMS.}
    \label{fig:global_histograms}
\end{figure*}

\begin{table*}
	\centering
	\caption{Volume-weighted mean of the light-weighted averages of $A_V$ dust attenuation affecting the stellar populations, $A_V$ dust attenuation derived from the Balmer decrement, and the ratio of the two quantities, for galaxies categorised (1) by stellar mass; (2) by $\rm sSFR_{H \alpha}$ relative to the SFMS, $\Delta {\rm log} \left(\Psi_{*, \: \rm H \alpha} / \rm{yr^{-1}} \right)$. The standard errors on the mean are quoted to $1 \sigma$. Also shown are the statistical significance levels between pairs of distributions in Fig.~\ref{fig:global_histograms}, determined via KS testing. Each sigma value corresponds to how significant the distinction is between a given distribution and the distribution in the table which immediately precedes it.}
	\label{tab:global_ratios}
	\begin{tabular}{lcccccc} 
		\hline
		Stellar mass & \multicolumn{2}{c}{Mean $A_{V, \: \rm stars}$} & \multicolumn{2}{c}{Mean $A_{V, \: \rm gas}$} & \multicolumn{2}{c}{Mean $A_{V, \: \rm stars} / A_{V, \: \rm gas}$}
		\vspace{1mm} \\
		 & Value & Significance & Value & Significance & Value & Significance\\
		\hline
		${\rm log} \left(M_{*} / \rm{M_{\odot}} \right) < 10.26$ & $0.27 \pm 0.02$ & --- & $0.62 \pm 0.03$ & --- & $0.43 \pm 0.03$ & ---\\
		$10.26 < {\rm log} \left(M_{*} / \rm{M_{\odot}} \right) < 10.58$ & $0.36 \pm 0.02$ & $2.0 \sigma$ & $0.99 \pm 0.04$ & $> 5 \sigma$ & $0.36 \pm 0.02$ & $2.3 \sigma$\\
		$10.58 < {\rm log} \left(M_{*} / \rm{M_{\odot}} \right) < 10.86$ & $0.33 \pm 0.02$ & $< 1 \sigma$ & $1.18 \pm 0.05$ & $3.4 \sigma$ & $0.29 \pm 0.02$ & $1.7 \sigma$\\
		${\rm log} \left(M_{*} / \rm{M_{\odot}} \right) > 10.86$ & $0.30 \pm 0.02$ & $< 1 \sigma$ & $1.20 \pm 0.03$ & $2.7 \sigma$ & $0.25 \pm 0.01$ & $2.9 \sigma$\\
		\hline
		$\rm sSFR_{H \alpha}$ relative to the SFMS & \multicolumn{2}{c}{Mean $A_{V, \: \rm stars}$} & \multicolumn{2}{c}{Mean $A_{V, \: \rm gas}$} & \multicolumn{2}{c}{Mean $A_{V, \: \rm stars} / A_{V, \: \rm gas}$}
		\vspace{1mm} \\
		 & Value & Significance & Value & Significance & Value & Significance\\
		\hline
		$\Delta {\rm log} \left(\Psi_{*, \: \rm H \alpha} / \rm{yr^{-1}} \right) < -0.71$ & $0.25 \pm 0.02$ & --- & $0.73 \pm 0.05$ & --- & $0.41 \pm 0.03$ & ---\\
		$-0.71 < \Delta {\rm log} \left(\Psi_{*, \: \rm H \alpha} / \rm{yr^{-1}} \right) < -0.39$ & $0.28 \pm 0.02$ & $< 1 \sigma$ & $0.83 \pm 0.04$ & $< 1 \sigma$ & $0.35 \pm 0.02$ & $< 1 \sigma$\\
		$-0.39 < \Delta {\rm log} \left(\Psi_{*, \: \rm H \alpha} / \rm{yr^{-1}} \right) < -0.10$ & $0.37 \pm 0.02$ & $3.0 \sigma$ & $0.97 \pm 0.05$ & $< 1 \sigma$ & $0.39 \pm 0.02$ & $2.1 \sigma$\\
		$\Delta {\rm log} \left(\Psi_{*, \: \rm H \alpha} / \rm{yr^{-1}} \right) > -0.10$ & $0.34 \pm 0.03$ & $3.7 \sigma$ & $0.92 \pm 0.05$ & $2.7 \sigma$ & $0.35 \pm 0.03$ & $2.0 \sigma$\\
		\hline
	\end{tabular}
\end{table*}

Figure~\ref{fig:global_histograms} shows the distributions of the global $A_V$ dust attenuation affecting the stellar populations (hereafter denoted $A_{V, \: \rm stars}$), global $A_V$ dust attenuation derived from the Balmer decrement (hereafter $A_{V, \: \rm gas}$), and the ratio of the two quantities (hereafter $A_{V, \: \rm stars} / A_{V, \: \rm gas}$) for each of the four stellar mass categories and the four categories of $\rm sSFR_{H \alpha}$ relative to the SFMS. The average $A_{V, \: \rm stars}$ and $A_{V, \: \rm gas}$ values for each galaxy are light-weighted using the $g$-band flux. The histogram bins have been volume-weighted. Table~\ref{tab:global_ratios} summarises the volume-weighted means of the light-weighted averages for the global $A_{V, \: \rm stars}$, $A_{V, \: \rm gas}$, and the ratio between the two quantities, for each of the four stellar mass categories and the four categories of $\rm sSFR_{H \alpha}$ relative to the SFMS.

Both Fig.~\ref{fig:global_histograms} and Table~\ref{tab:global_ratios} show that $A_{V, \: \rm gas}$ consistently has a higher mean value than $A_{V, \: \rm stars}$. This finding is in agreement with the work of those who have extensively studied dust within starburst galaxies \citep[e.g.][]{Fanelli1988SpectralGalaxies, Keel1993ObscurationNuclei, Calzetti1994DustLaw, Calzetti2000TheGalaxies, Kreckel2013a}. Both the peaks of the distributions in Fig.~\ref{fig:global_histograms} and the mean global $A_{V, \: \rm stars} / A_{V, \: \rm gas}$ ratios in Table~\ref{tab:global_ratios} may be compared with the value obtained by \citet{Calzetti2000TheGalaxies}, who found that $E(B-V)_{\rm stars} / E(B-V)_{\rm gas} = 0.44 \pm 0.03$. Accounting for the differing values of $R_V$ used in the reddening curves for the stellar populations and the gas in this work (see Section~\ref{subsec:Balmer Decrement}), this ratio of colour excesses corresponds to a ratio of $A_{V, \: \rm stars} / A_{V, \: \rm gas} = 0.57 \pm 0.04$ \citep{Calzetti2000TheGalaxies}. The ratios calculated in this work in Table~\ref{tab:global_ratios} are systematically lower than that found by \citet{Calzetti2000TheGalaxies}.

The results in this work, however, should not be directly compared with those of \citet{Calzetti2000TheGalaxies} for two main reasons. Firstly, the methods are different. \citet{Calzetti2000TheGalaxies} derived the dust attenuation affecting the stellar populations by measuring the shape of the observed UV continuum; in this work, however, these attenuation values are obtained through the use of the full-spectrum stellar population synthesis code \texttt{STARLIGHT} \citep{CidFernandes2005Semi-empiricalMethod}, which assumes a reddening curve for the stellar continuum. Secondly, the samples are different. The galaxies analysed by \citet{Calzetti2000TheGalaxies} were starburst galaxies with exceptionally hyperactive SFRs: the mean and median SFRs of their galaxies are $18 \pm 5 \: \rm M_{\odot} \ yr^{-1}$ and $5 \pm 6 \: \rm M_{\odot} \ yr^{-1}$ respectively \citep{Storchi-Bergmann1994UltravioletEffects}. By contrast, the mean and median $\rm H \alpha$ derived SFRs of the MaNGA galaxies, at $1.6 \pm 0.1 \: \rm M_{\odot} \ yr^{-1}$ and $1.0 \pm 0.2 \: \rm M_{\odot} \ yr^{-1}$ respectively, are more representative of normal SF spiral galaxies.

\medskip

We now consider the effect that both stellar mass and global $\rm sSFR_{H \alpha}$ relative to the main sequence of SF galaxies have upon both of the dust attenuation measures, and the ratio between them.

Figure~\ref{fig:global_histograms} shows that there is a dependence on the stellar mass of the galaxies for the global light-weighted $A_{V, \: \rm gas}$ and the $A_{V, \: \rm stars} / A_{V, \: \rm gas}$ ratio: the global light-weighted $A_{V, \: \rm gas}$ increases and the global light-weighted $A_{V, \: \rm stars} / A_{V, \: \rm gas}$ ratio decreases as the stellar mass of the galaxies increases. However, this effect is non-existent for the global light-weighted $A_{V, \: \rm stars}$, which remains roughly constant in each of the mass categories. Kolmogorov-Smirnov (KS) testing of the distributions in Fig.~\ref{fig:global_histograms} reveals that the dependence on stellar mass for the global light-weighted $A_{V, \: \rm gas}$ and the $A_{V, \: \rm stars} / A_{V, \: \rm gas}$ ratio is statistically significant, but finds no statistical difference between the distributions in the global light-weighted $A_{V, \: \rm stars}$. The mean values of all of these quantities for each stellar mass category, as well as the significance levels to which the distributions are distinct, are summarised in Table~\ref{tab:global_ratios}.

Further KS testing of the distributions in Fig.~\ref{fig:global_histograms} finds that there is only a weak dependence on global $\rm sSFR_{H \alpha}$ relative to the SFMS for each of the global light-weighted $A_{V, \: \rm stars}$ and $A_{V, \: \rm gas}$: as the global $\rm sSFR_{H \alpha}$ of the galaxies decreases relative to the SFMS, the values of both the attenuation measures decrease. No statistically significant dependence on global $\rm sSFR_{H \alpha}$ relative to the SFMS is detected for the ratio of these two quantities; Table~\ref{tab:global_ratios} shows that the mean values for the the global light-weighted $A_{V, \: \rm stars} / A_{V, \: \rm gas}$ ratio remain roughly constant at ${\sim} 0.35$.

\subsection{Looking Locally within the Galaxies}
\label{subsec:Looking Locally}

We now use the spatial resolution of MaNGA to look not just at the global properties of the galaxies, but their local properties as well. In particular, we investigate how dust attenuation depends on the location within galaxies: how it varies spatially, with galactocentric radius, and between the spiral arm and inter-arm regions. Moreover, we study how these dust attenuation properties vary between the stellar mass and $\rm sSFR_{H \alpha}$ categories defined in Section~\ref{subsec:Categorising galaxies}.

\subsubsection{Dust Attenuation within Spiral Arms is not Resolved}
\label{subsubsec:Spiral Arms}

Since dust lanes are often associated with the spiral arms of galaxies \citep[e.g.][]{Tamburro2008GeometricallyGalaxies, Hou2015OffsetWay, Shabani2018SearchGalaxies}, we use the GZ:3D spiral arm masks to decompose each of the galaxies in the sample into their spiral arm and inter-arm regions. Ultimately, however, almost no significant difference is seen between any of the dust attenuation quantities described so far for data inside the spiral arms and in the inter-arm regions. This null result is likely due to the relatively large FWHM of the MaNGA PSF of ${\sim} 2.5"$ \citep{Bundy2015OVERVIEWOBSERVATORY}; an angular distance of ${\sim} 2 \ \rm kpc$ at the average distance of the sample. At such a spatial resolution, the results are smeared and it is difficult to resolve whether dust lanes, which tend to be narrow \citep[e.g][]{Xilouris1999AreThick, Dalcanton2004TheEvolution, Holwerda2012NewNHEMESES}, are located in the spiral arm or inter-arm region of the GZ:3D mask. This effect is compounded by the fact that GZ:3D respondents would naturally select the brightest part of the image as the spiral arm, which also is likely to exclude dust lanes which lag behind spiral arms \citep[e.g.][]{Egusa2004OffsetsGalaxies, Chandar2017CluesClusters} from the GZ:3D masks employed in this work. Finally, dust lanes are optically thick \citep[e.g.][]{White1992DirectGalaxy, White2000SeeingGalaxies, Holwerda2004TheDisks}, which will result in regions of reduced brightness, meaning they might be intentionally omitted from the GZ:3D masks.

\begin{figure*}
\centering
	\includegraphics[width=0.26\textwidth]{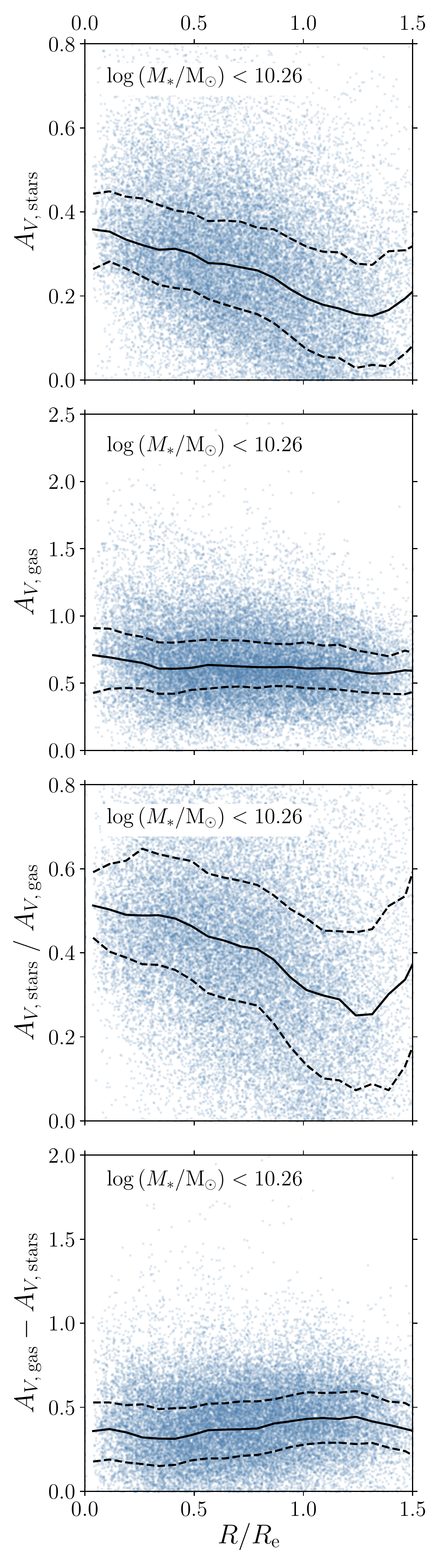}
	\includegraphics[width=0.23\textwidth]{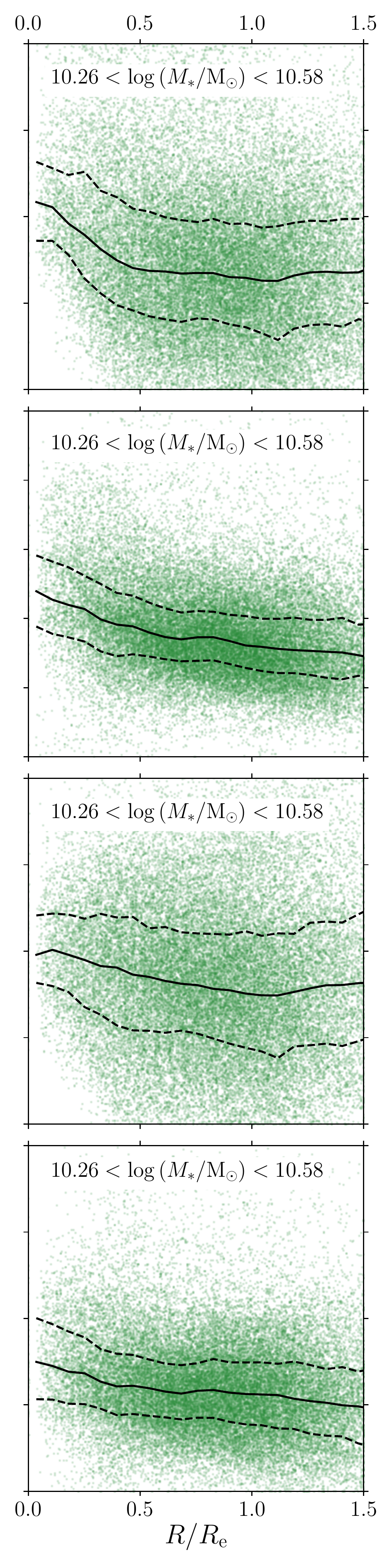}
	\includegraphics[width=0.23\textwidth]{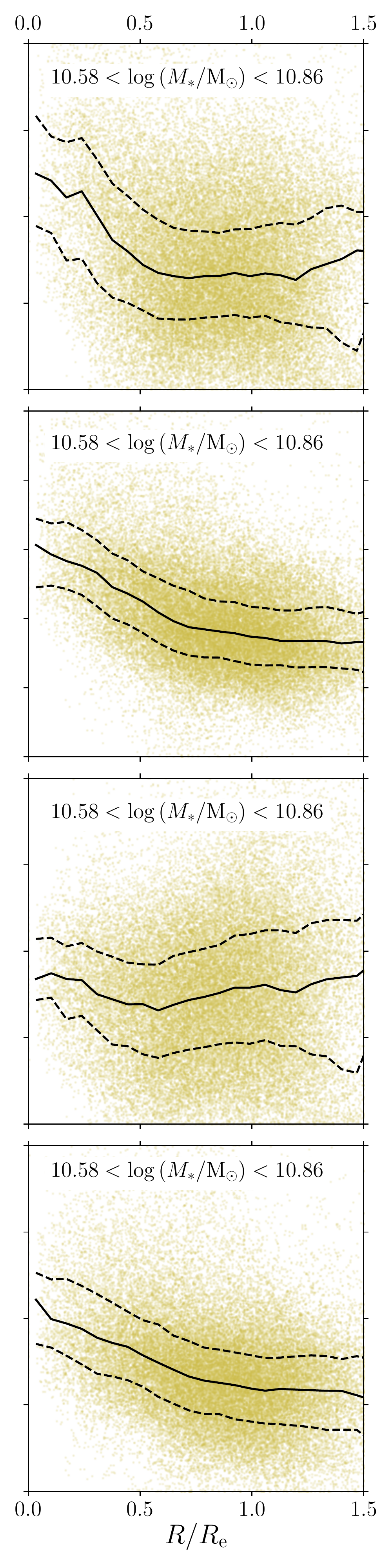}
	\includegraphics[width=0.23\textwidth]{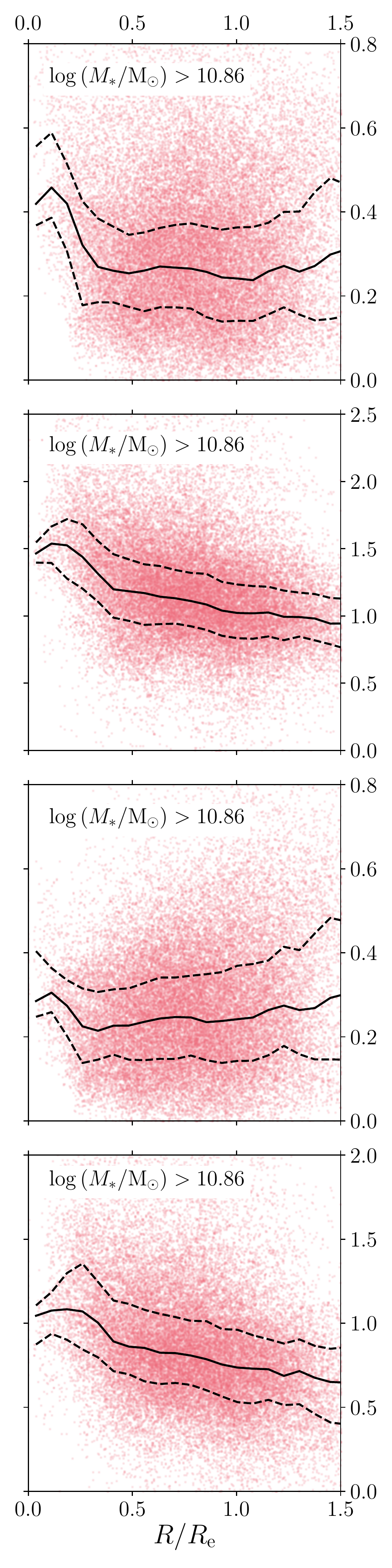}
    \caption{All spaxels plotted on scatter graphs showing the local $A_V$ dust attenuation affecting the stellar populations (top row), the local $A_V$ dust attenuation derived from the Balmer decrement (upper-middle row), the ratio of the two quantities (lower-middle row), and the difference between the two quantities (bottom row) as a function of distance from the galactic centre in units of $R_{\rm e}$. The blue points (far left column) are data from galaxies with stellar masses in the range ${\rm log} \left(M_{*} / \rm{M_{\odot}} \right) < 10.26$, green (centre-left column) data are from galaxies with $10.26 < {\rm log} \left(M_{*} / \rm{M_{\odot}} \right) < 10.58$, yellow (centre-right column) data are from galaxies with $10.58 < {\rm log} \left(M_{*} / \rm{M_{\odot}} \right) < 10.86$, and the red (far right column) data from galaxies with ${\rm log} \left(M_{*} / \rm{M_{\odot}} \right) > 10.86$. The running medians (solid black lines) and inter-quartile ranges (dashed black lines) to the data have also been added to each plot. Each of these running averages have been volume-weighted.}
    \label{fig:gradient_radius_graphs_mass}
\end{figure*}

\begin{figure*}
\centering
	\includegraphics[width=0.26\textwidth]{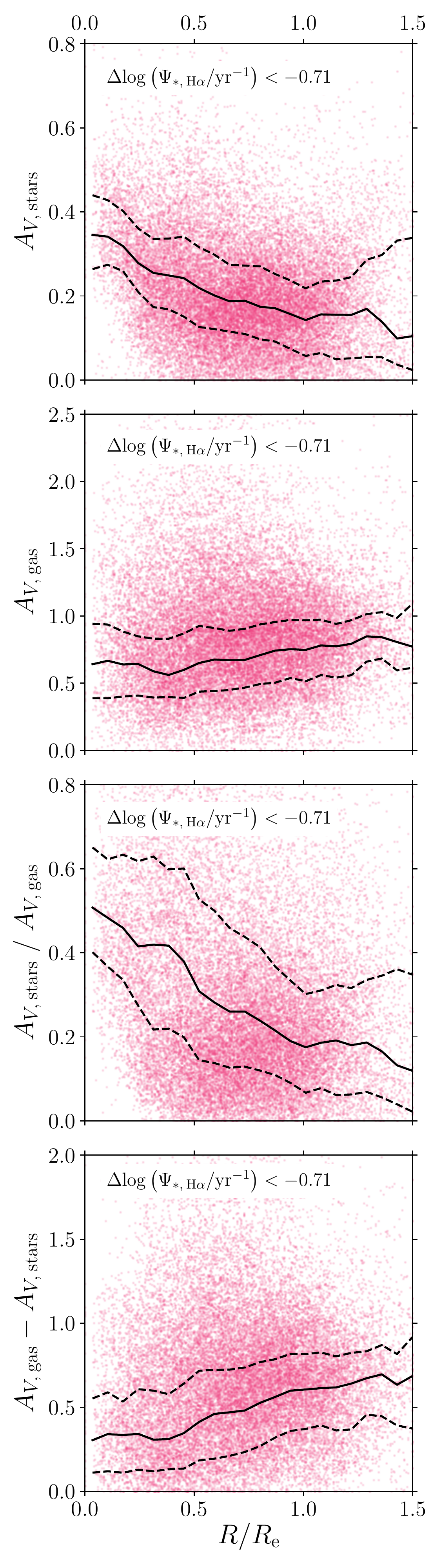}
	\includegraphics[width=0.23\textwidth]{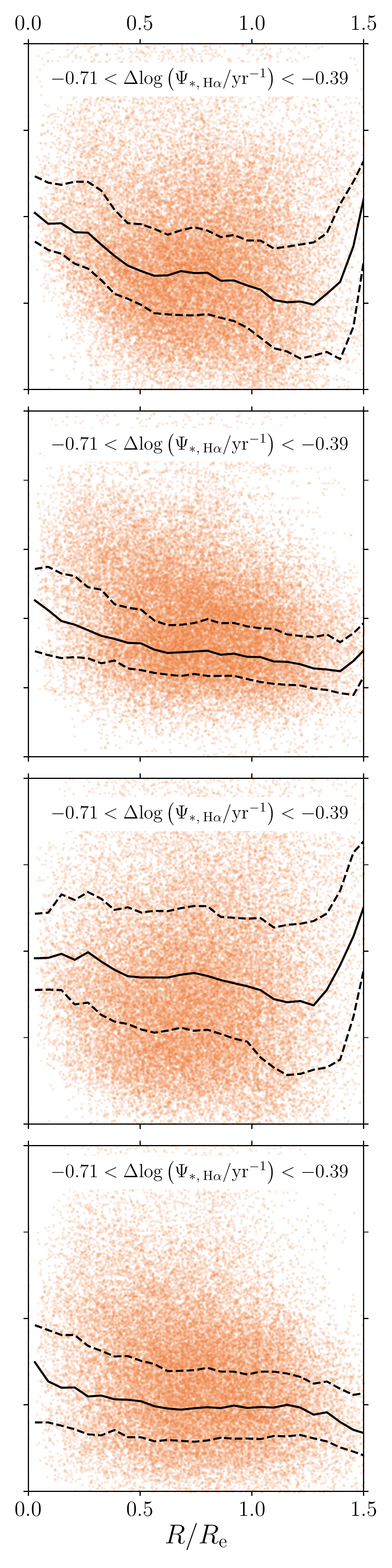}
	\includegraphics[width=0.23\textwidth]{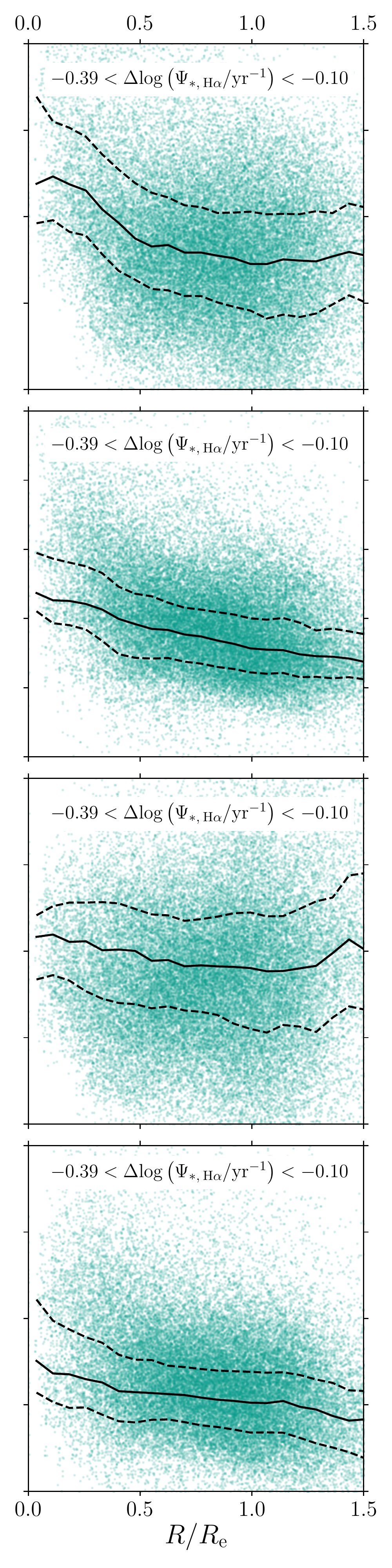}
	\includegraphics[width=0.23\textwidth]{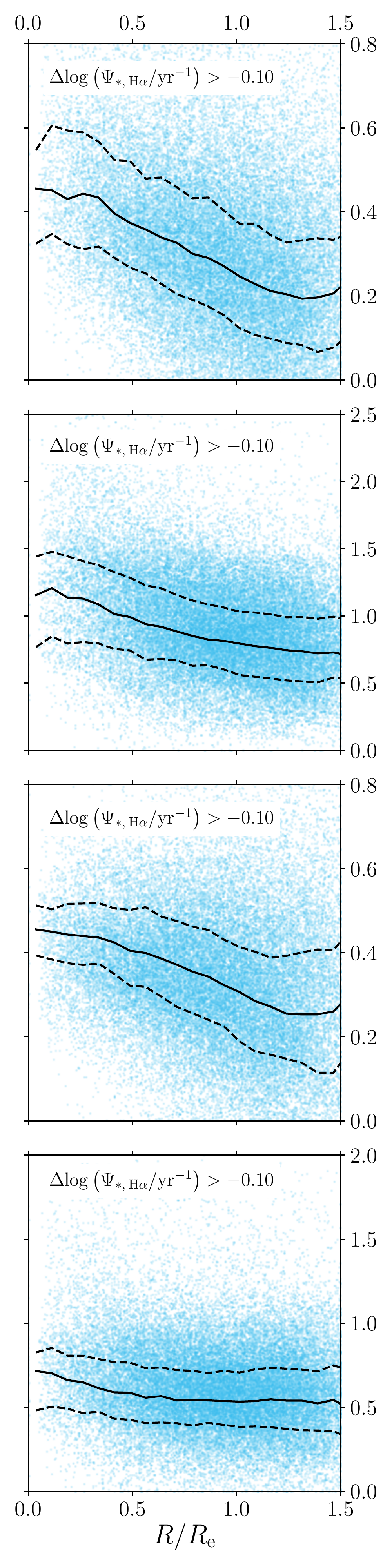}
    \caption{As in Fig.~\ref{fig:gradient_radius_graphs_mass}, except galaxies are now categorised by their global $\rm sSFR_{H \alpha}$ relative to the SFMS. The magenta points (far left column) are data from galaxies with a $\rm sSFR_{H \alpha}$ of $\Delta {\rm log} \left(\Psi_{*, \: \rm H \alpha} / \rm{yr^{-1}} \right) < -0.71$ from the SFMS, orange (centre-left column) data are from galaxies with a $\rm sSFR_{H \alpha}$ of $-0.71 < \Delta {\rm log} \left(\Psi_{*, \: \rm H \alpha} / \rm{yr^{-1}} \right) < -0.39$ from the SFMS, teal (centre-right column) data are from galaxies with a $\rm sSFR_{H \alpha}$ of $-0.39 < \Delta {\rm log} \left(\Psi_{*, \: \rm H \alpha} / \rm{yr^{-1}} \right) < -0.10$ from the SFMS, and the cyan (far right column) data are from galaxies with a $\rm sSFR_{H \alpha}$ of $\Delta {\rm log} \left(\Psi_{*, \: \rm H \alpha} / \rm{yr^{-1}} \right) > -0.10$ from the SFMS.}
    \label{fig:gradient_radius_graphs_sSFR}
\end{figure*}

\subsubsection{Average and Radial Variation of Dust Attenuation: Dependence on Stellar Mass}
\label{subsubsec:Radial Dependence Stellar Mass}

We next investigate how the average dust attenuation quantities depend on galaxy stellar mass. Figure~\ref{fig:gradient_radius_graphs_mass} shows graphs of each of local $A_V$ dust attenuation affecting the stellar populations, local $A_V$ dust attenuation derived from the Balmer decrement, and finally the ratio of, and the difference between, these two quantities as a function of radius. The blue points (far left column) are data from the lowest-mass galaxies in the sample. Moving rightwards, the stellar mass increases; the red data (far right column) are from the highest-mass galaxies in the sample. Volume-weighted running medians and inter-quartile ranges are overlaid on the data.

Figure~\ref{fig:gradient_radius_graphs_mass} compares how the median local dust attenuation affecting the stellar populations, $A_{V, \: \rm stars}$, and the median local dust attenuation derived from measurements of the Balmer decrement, $A_{V, \: \rm gas}$, the ratio between the two attenuation measures, and the $A_{V, \: \rm gas}$ excess are affected by stellar mass. Of these quantities, the median local $A_{V, \: \rm gas}$ depends most strongly on stellar mass: its average value increases for increasing stellar mass. This finding is in agreement with those of other authors, who report higher average values for the local $A_{V, \: \rm gas}$ for higher-mass galaxies \citep{Nelson2015Spatially-resolvedZ1.4, Jafariyazani2019SpatiallyQuenching}. By contrast, the median local $A_{V, \: \rm stars}$ does not change appreciably between each stellar mass category (although see \citealp{GonzalezDelgado2015TheSequence} and \citealp{Goddard2016SDSS-IVType}, both of whom do find higher average values for $A_{V, \: \rm stars}$ in galaxies with higher stellar masses. We likely do not see this effect since these authors had at their disposal a much wider range of stellar masses than in this work). At any rate, this work shows that the average value of $A_{V, \: \rm gas}$ is more strongly influenced by stellar mass than the average $A_{V, \: \rm stars}$; this in turn means that the average $A_{V, \: \rm stars} / A_{V, \: \rm gas}$ ratio decreases for increasing stellar mass.

\medskip

So far, we have determined the $A_{V, \: \rm stars} / A_{V, \: \rm gas}$ ratio for the sake of comparison with the work of \citet{Calzetti2000TheGalaxies}. Although this ratio is a useful diagnostic for determining global properties of galaxies, it is actually not physically motivated, since $A_{\lambda}$ is a logarithmic quantity measured in magnitudes \citep[see for instance][]{Salim2020TheGalaxies}. Consequently, the ratio of $A_{V, \: \rm stars}$ to $A_{V, \: \rm gas}$ -- a ratio of two logarithms -- does not represent a physically meaningful quantity.

We therefore here introduce a new quantity, which we coin the $A_{V, \: \rm gas}$ excess, defined as $A_{V, \: \rm gas} - A_{V, \: \rm stars}$, and which unlike the $A_{V, \: \rm stars} / A_{V, \: \rm gas}$ ratio is physically motivated. Since (as we discuss in Section~\ref{subsec:Stellar Populations}) \texttt{STARLIGHT} provides little information about the dust attenuation affecting the youngest stellar populations, $A_{V, \: \rm stars}$ primarily quantifies the contribution to the dust attenuation from the diffuse ISM. By contrast, $A_{V, \: \rm gas}$ quantifies the dust attenuation due to the combination of this diffuse ISM and the clumpy BCs. Therefore, by subtracting $A_{V, \: \rm stars}$ from $A_{V, \: \rm gas}$ it is possible to estimate the dust attenuation \emph{due the BCs alone}. Note that the $A_{V, \: \rm gas}$ excess probably over-estimates the dust attenuation due to the BCs, since \texttt{STARLIGHT} calculates the dust attenuation affecting the stellar populations down to ${\sim} 3 \times 10^7 \rm yr$, which are relatively young and will hence experience both attenuation from the diffuse ISM and their BCs.

Since measurements of the $A_{V, \: \rm gas}$ excess can also provide information about local variations in the dust geometry of SF galaxies, we now examine how the average of this newly-defined quantity varies with stellar mass. Figure~\ref{fig:gradient_radius_graphs_mass} shows that a similar (though less pronounced) effect is seen in both the variation of the average $A_{V, \: \rm gas}$ excess and the average $A_{V, \: \rm gas}$, both of which increase as stellar mass increases.

\medskip

As well as showing the average values of the dust attenuation quantities, Fig.~\ref{fig:gradient_radius_graphs_mass} also shows that both the local $A_{V, \: \rm stars}$ and local $A_{V, \: \rm gas}$ exhibit some radial dependence: the inner regions possess higher values for both dust attenuation quantities, which flatten out beyond ${\sim} 0.5 \ R_{\rm e}$. Each stellar mass category sees the value of $A_{V, \: \rm stars}$ fall from around ${\sim} 0.4-0.5 \: \rm{mag}$ at the galactic centre to ${\sim} 0.2-0.3 \: \rm{mag}$ at $1.5 \ R_{\rm e}$. Furthermore, the lowest-mass galaxies see no radial variation in their local $A_{V, \: \rm gas}$, which remains constant at around ${\sim} 0.6 \: \rm{mag}$, whereas for the highest-mass galaxies the local $A_{V, \: \rm gas}$ falls almost a magnitude from a value of ${\sim} 1.6 \: \rm{mag}$ to ${\sim} 0.8 \: \rm{mag}$. It is tempting to explain away this lack of radial variation in the local $A_{V, \: \rm gas}$ for the lowest-mass galaxies by assuming that the bulges of low-mass galaxies are proportionally smaller than high-mass galaxies, thereby causing galaxies with lower stellar masses to exhibit lower values of local $A_{V, \: \rm gas}$ nearer to their central regions. However, analysis of the bulge-to-total (B/T) light ratio of each galaxy (see \citealp{Simard2011ASurvey}) reveals that the mean B/T ratio is broadly similar for each category: this difference, therefore, is likely not a consequence of low-mass galaxies having small bulges in proportion to their size.

These findings are in agreement with previous work: both \citet{GonzalezDelgado2015TheSequence} as well as \citet{Goddard2016SDSS-IVType} find that dust attenuation affecting the stellar populations exhibits a similar radial gradient for both CALIFA survey galaxies and MaNGA galaxies, respectively. Furthermore, both \citet{Nelson2015Spatially-resolvedZ1.4} and \citet{Jafariyazani2019SpatiallyQuenching} show that dust attenuation derived from measurements of the Balmer decrement declines radially in a very similar manner for both 3D-HST survey galaxies and MUSE-Wide survey galaxies, respectively. Moreover, these authors also report steeper radial gradients for higher-mass galaxies for both the local $A_{V, \: \rm stars}$ \citep{GonzalezDelgado2015TheSequence, Goddard2016SDSS-IVType} and for the local $A_{V, \: \rm gas}$ \citep{Nelson2015Spatially-resolvedZ1.4, Jafariyazani2019SpatiallyQuenching}. Again, this behaviour can be seen in the plots of Fig.~\ref{fig:gradient_radius_graphs_mass}.

The local $A_{V, \: \rm stars} / A_{V, \: \rm gas}$ ratio is not particularly affected by radius at any galaxy stellar mass. The highest-mass galaxies (red) exhibit only a very small gradient close to their centres, after which $A_{V, \: \rm stars} / A_{V, \: \rm gas}$ remains reasonably constant with radius; the ratio for the intermediate-mass galaxies (yellow and green) is roughly constant between the galactic centre and $1.5 \ R_{\rm e}$. This behaviour is intriguing considering that $A_{V, \: \rm stars}$ and $A_{V, \: \rm gas}$ are independent measures of dust attenuation: to yield a ratio that is roughly constant with radius, both the local $A_{V, \: \rm stars}$ and $A_{V, \: \rm gas}$ must decrease with radius at approximately the same rate. The only exception is the local $A_{V, \: \rm stars} / A_{V, \: \rm gas}$ for the lowest-mass galaxies (blue), which exhibit a stronger radial gradient for this ratio, driven by the lack of such a gradient in the local $A_{V, \: \rm gas}$. Quite why the lowest-mass galaxies exhibit this behaviour is, however, unclear; it is possible that this could be a consequence of higher-mass galaxies having clumpier dust distributions and a greater amount of clumpy dust nearer to their centres than lower-mass galaxies.

The $A_{V, \: \rm gas}$ excess varies with radius in a very similar manner to the local $A_{V, \: \rm gas}$. The $A_{V, \: \rm gas}$ excess is higher at the centres of the galaxies than at their outskirts (again with the exception of those galaxies with the lowest stellar masses). This behaviour suggests that the radial gradients seen in the local $A_{V, \: \rm gas}$, both in this work and in other work \citep[e.g.][]{Nelson2015Spatially-resolvedZ1.4, Jafariyazani2019SpatiallyQuenching} cannot be attributed to a radial gradient in the diffuse ISM alone.

\subsubsection{Average and Radial Variation of Dust Attenuation: Dependence on Relative Global sSFR}
\label{subsubsec:Radial Dependence sSFR}

When the same analysis is carried out based on the categories of relative global $\rm sSFR_{H \alpha}$, similar results are obtained. These results are shown in Fig.~\ref{fig:gradient_radius_graphs_sSFR}, which plots each of the four dust attenuation quantities described in Section~\ref{subsubsec:Radial Dependence Stellar Mass} as a function of radius. The magenta data (far left column) are from the galaxies with the lowest relative global $\rm H \alpha$ derived sSFRs. Moving rightwards, the relative global $\rm sSFR_{H \alpha}$ increases; the cyan data (far right column) are from the galaxies with the highest relative global $\rm H \alpha$ derived sSFRs.

The median values of the local $A_{V, \: \rm stars}$, $A_{V, \: \rm gas}$, and the $A_{V, \: \rm gas}$ excess are less dependent on relative global $\rm sSFR_{H \alpha}$ than stellar mass. Nevertheless, all of these attenuation quantities have higher values for galaxies with high relative global $\rm H \alpha$ derived sSFRs compared with those with low relative global $\rm H \alpha$ derived sSFRs. In contrast to the results of Fig.~\ref{fig:gradient_radius_graphs_mass}, the average $A_{V, \: \rm stars} / A_{V, \: \rm gas}$ ratio is not dependent on relative global $\rm sSFR_{H \alpha}$; it has a constant median value of approximately ${\sim} 0.35$, since the average values of the local $A_{V, \: \rm stars}$ and $A_{V, \: \rm gas}$ are affected in very similar ways by the relative global $\rm sSFR_{H \alpha}$ of the galaxy. This contrast suggests that stellar mass has more influence on the local dust attenuation properties of a galaxy than the relative global $\rm sSFR_{H \alpha}$.

Strong radial gradients are found for the local $A_{V, \: \rm stars}$ for each of the four relative global $\rm sSFR_{H \alpha}$ categories, as well as for the local $A_{V, \: \rm gas}$ for each of the categories (except the galaxies with the lowest relative global $\rm H \alpha$ derived sSFRs). The similar radial variation of these two independent measures of dust attenuation causes the local $A_{V, \: \rm stars} / A_{V, \: \rm gas}$ ratios to remain broadly constant between $0 - 1.5 \ R_{\rm e}$ for each of the categories (except for galaxies with the lowest relative global $\rm H \alpha$ derived sSFRs). The radial gradients in the $A_{V, \: \rm gas}$ excess in Fig.~\ref{fig:gradient_radius_graphs_sSFR}, however, are not as strong as those seen in Fig.~\ref{fig:gradient_radius_graphs_mass}. While this behaviour again suggests that both the diffuse ISM and the clumpy BCs increase in density nearer to the centres of the galaxies (apart from those with the lowest relative global $\rm H \alpha$ derived sSFRs), it also reiterates that the galaxy stellar mass is a more important driver of local dust attenuation properties than the relative global $\rm sSFR_{H \alpha}$.

\subsection{Variations in Dust Attenuation Properties are Driven Predominantly by Local Physics}
\label{subsec:Local Physics Drives}

\begin{figure}
	\includegraphics[width=0.5\textwidth]{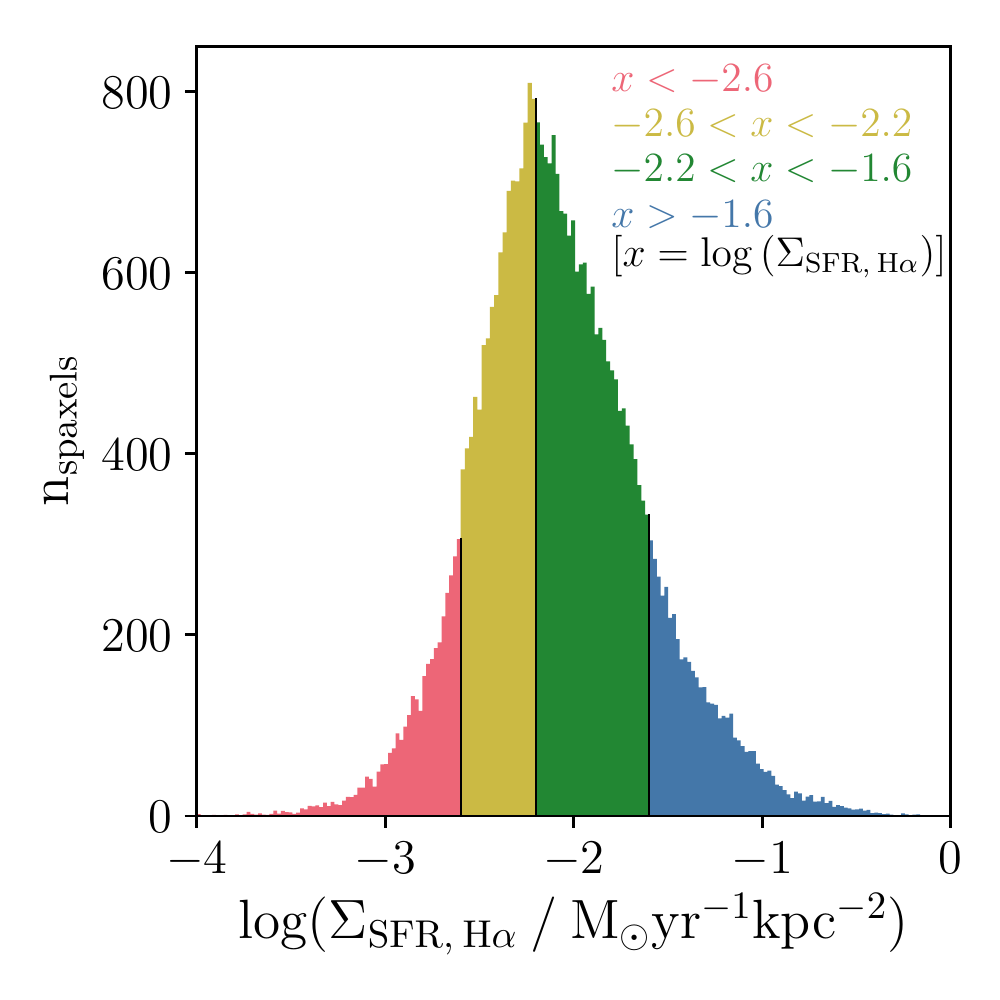}
    \caption{Histogram of all spaxels in the sample binned by $\rm SFR_{H \alpha}$ per unit physical surface area, $\Sigma_{\rm SFR, \: \rm H \alpha}$. The bins have been volume-weighted. The red region contains spaxels with the lowest $\Sigma_{\rm SFR, \: \rm H \alpha}$, with $\rm{log(\Sigma_{\rm SFR, \: \rm H \alpha} \: / \: \rm M_{\odot} \ yr^{-1} \ kpc^{-2})} < -2.6$ ; yellow contains $-2.6 < \rm{log(\Sigma_{\rm SFR, \: \rm H \alpha} \: / \: \rm M_{\odot} \ yr^{-1} \ kpc^{-2})} < -2.2$; green contains $-2.2 < \rm{log(\Sigma_{\rm SFR, \: \rm H \alpha} \: / \: \rm M_{\odot} \ yr^{-1} \ kpc^{-2})} < -1.6$; and the blue region contains spaxels with the highest $\Sigma_{\rm SFR, \: \rm H \alpha}$, with $\rm{log(\Sigma_{\rm SFR, \: \rm H \alpha} \: / \: \rm M_{\odot} \ yr^{-1} \ kpc^{-2})} > -1.6$.}
    \label{fig:SFR_SA}
\end{figure}

Having studied the effect of global galaxy properties on dust attenuation, we now consider the effect of local properties. The spatial resolution of MaNGA means that we do not have to group spaxels belonging to the same galaxies into the same categories: we can split the spaxels of each galaxy into multiple groups. In other words, the local effects of dust attenuation are now analysed, rather than purely global effects. Spaxels are categorised based on their $\rm SFR_{H \alpha}$ per unit physical surface area (hereafter denoted $\Sigma_{\rm SFR, \: \rm H \alpha}$), as seen in Fig.~\ref{fig:SFR_SA}. The histogram bins have been volume-weighted.

We compare the effect of global $\rm sSFR_{H \alpha}$ with local $\Sigma_{\rm SFR, \: \rm H \alpha}$ rather than with local $\rm sSFR_{H \alpha}$ since $\Sigma_{\rm SFR}$ is proportional to SFR per unit physical volume for disks of approximately uniform thickness. SFR per unit physical surface area is hence a more direct physical measurement of the star formation activity in a given region of a galaxy than SFR per unit stellar mass; moreover, local $\Sigma_{\rm SFR}$ is a less model-dependent measure of star formation activity than local sSFR and will still show similar results \citep[e.g.][]{GonzalezDelgado2016StarGalaxies, Medling2018TheFormation, Sanchez2019Spatially-ResolvedGalaxies}.

\medskip

\begin{figure*}
\centering
	\includegraphics[width=0.24\textwidth]{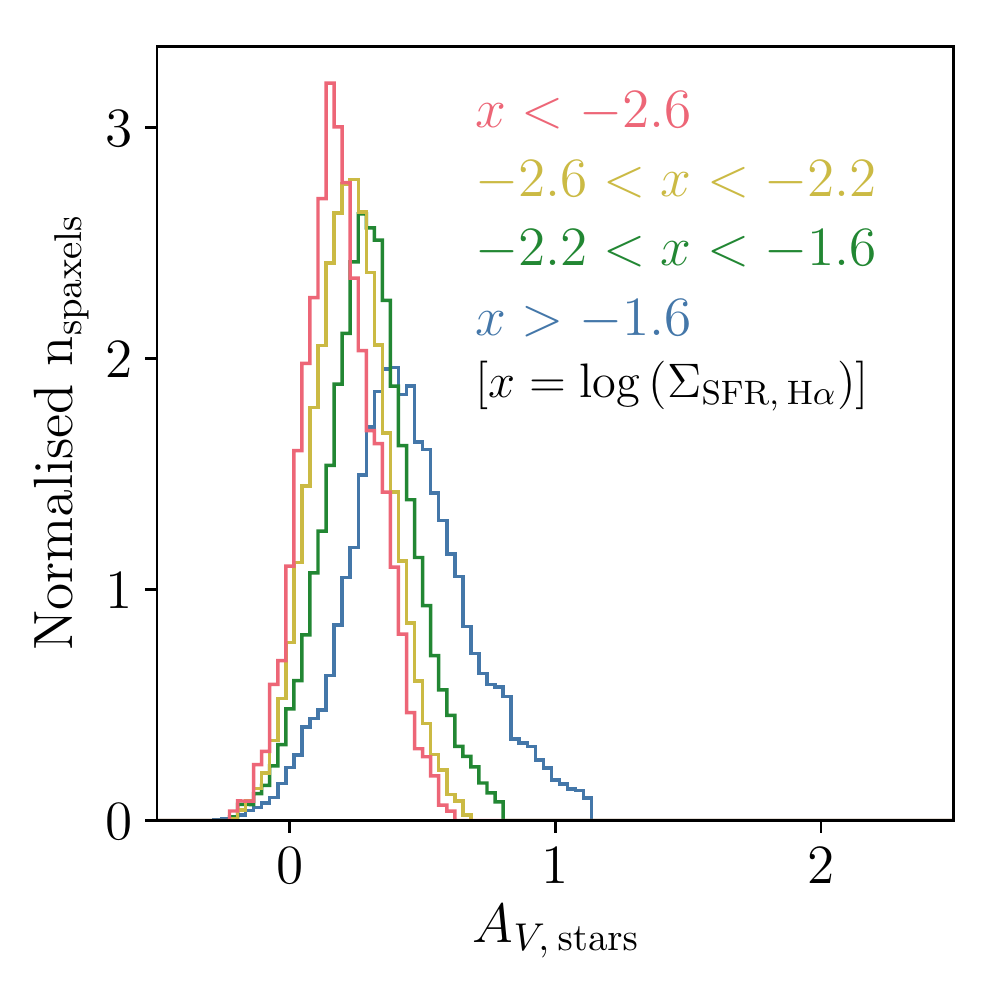}
	\includegraphics[width=0.24\textwidth]{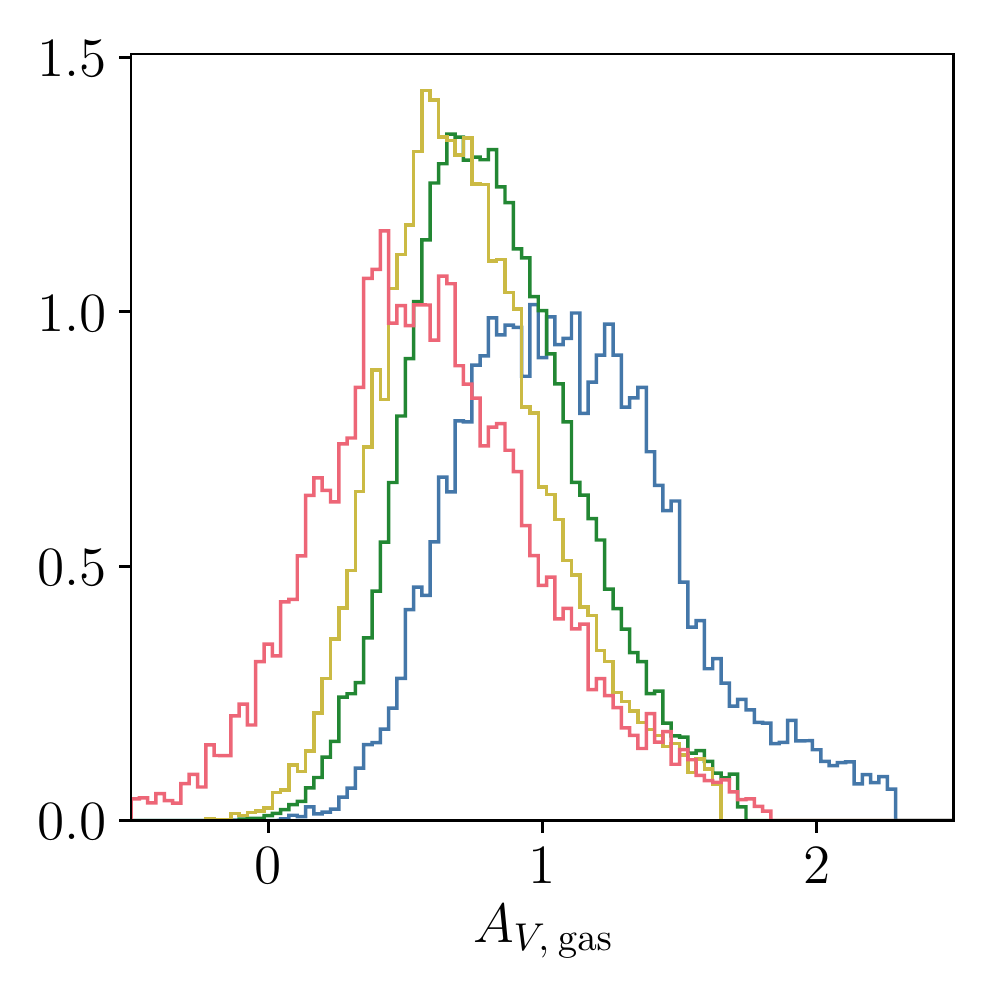}
	\includegraphics[width=0.24\textwidth]{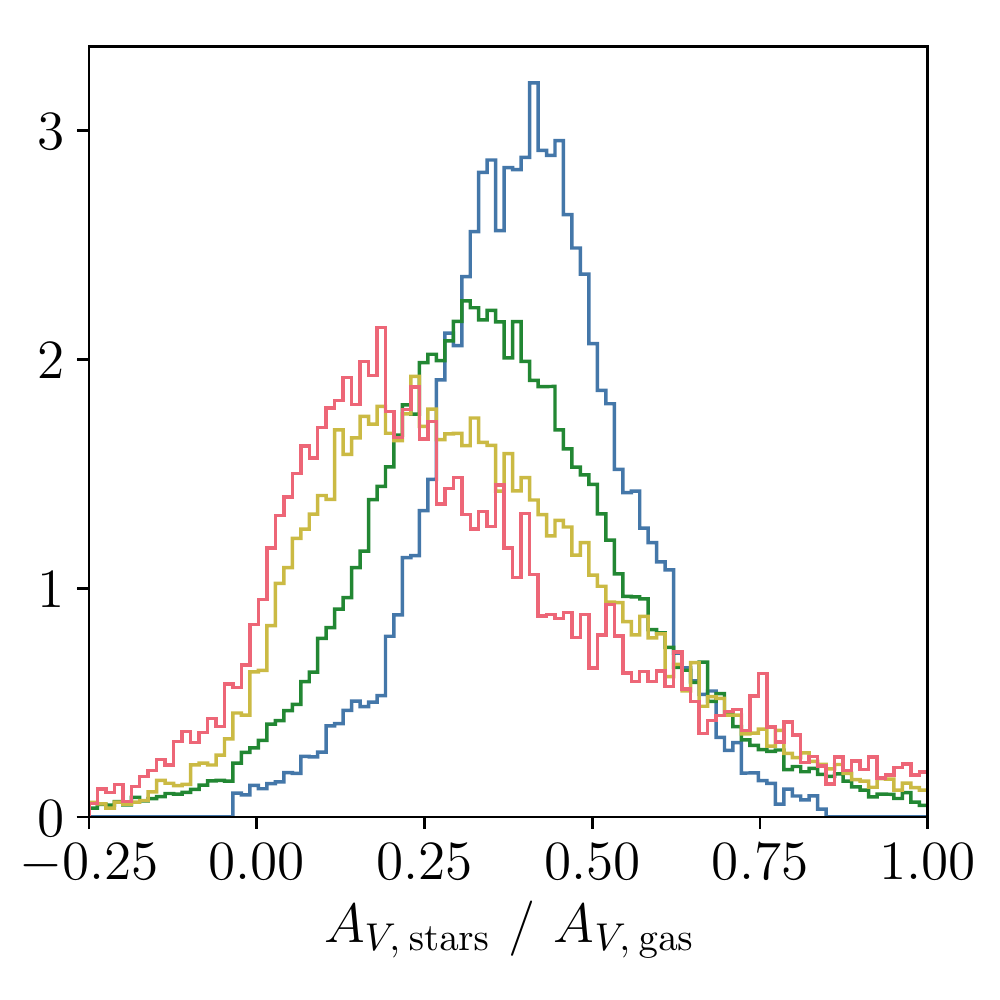}
	\includegraphics[width=0.24\textwidth]{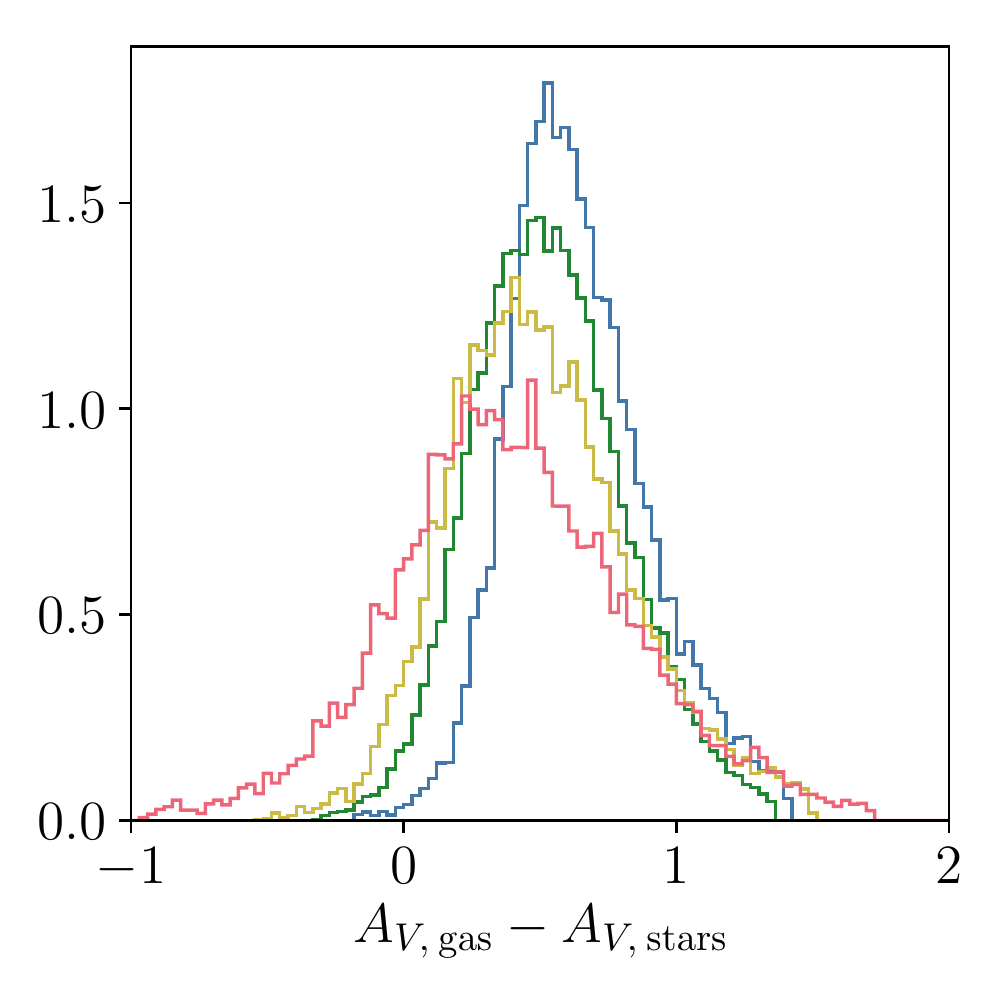}
    \caption{Distributions of (from left to right) $A_V$ dust attenuation affecting the stellar populations, $A_V$ dust attenuation derived from the Balmer decrement, the ratio of the two quantities, and the difference between the two quantities. The bins have been volume-weighted, and are normalised so that the area under each line is equal to one. The red histograms comprise spaxels with the lowest $\Sigma_{\rm SFR, \: \rm H \alpha}$, in the range $\rm{log(\Sigma_{\rm SFR, \: \rm H \alpha} \: / \: \rm M_{\odot} \ yr^{-1} \ kpc^{-2})} < -2.6$; yellow have $-2.6 < \rm{log(\Sigma_{\rm SFR, \: \rm H \alpha} \: / \: \rm M_{\odot} \ yr^{-1} \ kpc^{-2})} < -2.2$; green have $-2.2 < \rm{log(\Sigma_{\rm SFR, \: \rm H \alpha} \: / \: \rm M_{\odot} \ yr^{-1} \ kpc^{-2})} < -1.6$; and blue have the highest $\Sigma_{\rm SFR, \: \rm H \alpha}$, in the range $\rm{log(\Sigma_{\rm SFR, \: \rm H \alpha} \: / \: \rm M_{\odot} \ yr^{-1} \ kpc^{-2})} > -1.6$.}
    \label{fig:local_histograms}
\end{figure*}

Figure~\ref{fig:local_histograms} shows the distributions of the local $A_V$ dust attenuation affecting the stellar populations, local $A_V$ dust attenuation derived from the Balmer decrement, the ratio of the two quantities, and the difference between the two quantities. The histogram bins have been volume-weighted. Table~\ref{tab:local_ratios} summarises the mean values of the volume-weighted local $A_{V, \: \rm stars}$, the volume-weighted local $A_{V, \: \rm gas}$, and both the ratio and the difference between the two quantities for each of the four newly defined categories.

Figure~\ref{fig:local_histograms} shows that there is a strong dependence on $\Sigma_{\rm SFR, \: \rm H \alpha}$ for each of the local $A_{V, \: \rm stars}$, the local $A_{V, \: \rm gas}$, the ratio of these two quantities, and the local $A_{V, \: \rm gas}$ excess ($A_{V, \: \rm gas} - A_{V, \: \rm stars}$). KS tests confirm what can be seen by eye -- each of the pairs of neighbouring distributions in Fig.~\ref{fig:local_histograms} are found to be distinct to well over $5 \sigma$ confidence. As the $\Sigma_{\rm SFR, \: \rm H \alpha}$ of the spaxels increases, higher values are seen both for the dust attenuation affecting the stellar populations and the dust attenuation derived from measurements of the Balmer decrement. These trends are in agreement with previous work: SFR is shown to be strongly correlated with the total dust content of galaxies \citep[e.g.][]{DaCunha2008, daCunha2010NewGalaxies, Rowlands2014Herschel-ATLAS:Redshifts, Remy-Ruyer2015LinkingPicture}. Furthermore, SFR has been found to increase both as the dust attenuation affecting the stellar populations increases \citep[e.g][]{Tress2018SHARDS:Z2}, and as the dust attenuation derived from measurements of the Balmer decrement increases \citep[e.g][]{Garn2010PredictingGalaxy, Zahid2012TheGalaxies, Zahid2017StellarRate}. However, the studies undertaken by these authors analysed only the global properties of galaxies; the advantage of this work is to demonstrate that local $\Sigma_{\rm SFR, \: \rm H \alpha}$ variations have more influence than purely global quantities on dust attenuation properties.

\begin{table*}
	\centering
	\caption{Mean volume-weighted local $A_V$ dust attenuation affecting the stellar populations, $A_V$ dust attenuation derived from the Balmer decrement, the ratio of the two quantities, and the difference of the two quantities for each of the re-defined galaxy categories. The standard errors on the mean are quoted to $1 \sigma$.}
	\label{tab:local_ratios}
	\begin{tabular}{lcccc} 
		\hline
		$\rm{log(\Sigma_{\rm SFR, \: \rm H \alpha})} \: [ \rm M_{\odot} \ yr^{-1} \ kpc^{-2} ]$ & Mean $A_{V, \: \rm stars}$ & Mean $A_{V, \: \rm gas}$ & Mean $A_{V, \: \rm stars} / A_{V, \: \rm gas}$ & Mean $A_{V, \: \rm gas} - A_{V, \: \rm stars}$
		\vspace{1mm} \\
		\hline
		$\rm{log(\Sigma_{\rm SFR, \: \rm H \alpha})} < -2.6$ & $0.191 \pm 0.001$ & $0.592 \pm 0.004$ & $0.335 \pm 0.005$ & $0.400 \pm 0.004$\\
		$-2.6 < \rm{log(\Sigma_{\rm SFR, \: \rm H \alpha})} < -2.2$ & $0.2312 \pm 0.0007$ & $0.733 \pm 0.002$ & $0.338 \pm 0.001$ & $0.502 \pm 0.002$\\
		$-2.2 < \rm{log(\Sigma_{\rm SFR, \: \rm H \alpha})} < -1.6$ & $0.3000 \pm 0.0007$ & $0.833 \pm 0.001$ & $0.3635 \pm 0.0008$ & $0.534 \pm 0.001$\\
		$\rm{log(\Sigma_{\rm SFR, \: \rm H \alpha})} > -1.6$ & $0.454 \pm 0.002$ & $1.108 \pm 0.003$ & $0.405 \pm 0.001$ & $0.643 \pm 0.002$\\
		\hline
	\end{tabular}
\end{table*}

The local $A_{V, \: \rm gas}$ excess also increases for increasing $\Sigma_{\rm SFR, \: \rm H \alpha}$, which implies that the dependence of the local $A_{V, \: \rm gas}$ is not dominated by the dust attenuation due to the diffuse ISM alone. Finally, the $A_{V, \: \rm stars} / A_{V, \: \rm gas}$ ratio also increases as the $\Sigma_{\rm SFR, \: \rm H \alpha}$ of the spaxels increases, despite the fact that both the numerator and denominator of this ratio increase with increasing $\Sigma_{\rm SFR, \: \rm H \alpha}$. The other plots of Fig.~\ref{fig:local_histograms} demonstrate why this is the case: while the dependence on $\Sigma_{\rm SFR, \: \rm H \alpha}$ is strong for $A_{V, \: \rm gas}$, this dependence is stronger still for $A_{V, \: \rm stars}$. The $A_{V, \: \rm stars} / A_{V, \: \rm gas}$ ratio therefore increases for increasing $\Sigma_{\rm SFR, \: \rm H \alpha}$. This result is in contrast to that found when considering the purely global properties of the galaxies in Fig.~\ref{fig:global_histograms}.

The results obtained from Fig.~\ref{fig:local_histograms} provide a cautionary tale against putting too much stock in the mean global light- and volume-weighted quantities in Table~\ref{tab:global_ratios} and the mean local volume-weighted quantities in Table~\ref{tab:local_ratios}. The mean local $A_{V, \: \rm stars} / A_{V, \: \rm gas}$ ratios in Table~\ref{tab:local_ratios} especially are at odds with the distributions seen in Fig.~\ref{fig:local_histograms}, the peaks of which are distinct. For this reason, it is much more illuminating to interpret the results from the distributions of Fig.~\ref{fig:global_histograms} and Fig.~\ref{fig:local_histograms}, rather than simply the mean values in Table~\ref{tab:global_ratios} and Table~\ref{tab:local_ratios}.

The disparity between Fig.~\ref{fig:global_histograms} and Fig.~\ref{fig:local_histograms} implies that variations in dust attenuation properties are most likely driven more by local physics than by global physics. Locally, $A_{V, \: \rm gas}$ is strongly dependent on $\Sigma_{\rm SFR, \: \rm H \alpha}$ because by isolating only those spaxels within galaxies with higher $\Sigma_{\rm SFR, \: \rm H \alpha}$, a greater number of clumpy BCs associated with ongoing star formation should be expected. By contrast, when considering the galaxies globally this effect is washed out because each galaxy is likely to have regions of high $\rm sSFR_{H \alpha}$ balanced out by regions where the local $\rm sSFR_{H \alpha}$ is lower.

The difference between the global and local distributions for the dust attenuation affecting the stellar populations is not as pronounced. The peaks of the two separate sets of distributions are much closer together than those for the $A_{V, \: \rm gas}$ are. The distributions for $A_{V, \: \rm stars}$ in Fig.~\ref{fig:global_histograms} and Fig.~\ref{fig:local_histograms} will average over the clumpy dust since some stars will be in front of the diffuse, dusty ISM (experiencing little attenuation) and some will be behind. This means that there is not as much difference between the global and local $A_{V, \: \rm stars}$ distributions. In short, the results of Fig.~\ref{fig:local_histograms} imply that while variations in the $A_{V, \: \rm stars}$ are probably a result of both local and global physics, variations in the $A_{V, \: \rm gas}$, by contrast, are predominantly driven by local physics.

\section{The Geometry of Dust in Star-Forming Spiral Galaxies}
\label{sec:Dust Geometry}

Let us now turn to an interpretation of these results. Models for the geometry of dust in SF galaxies -- such as that depicted in Fig.~13 of \citet{Wild2011a} -- generally incorporate four key components:

\begin{enumerate}
  \item \emph{Diffuse, dusty ISM:} Exhibits a radial gradient -- the density of the ISM decreases with galactocentric radius.
  
  \vspace{1mm}
  
  \item \emph{Young stars:} These highly ionising O and B stars are surrounded by \textsc{H~ii} regions, and thus produce nebular emission lines. These stars are located in the disk, and are still closely associated with their dusty, clumpy BCs.
  
  \vspace{1mm}
  
  \item \emph{Intermediate age stars:} Also found in the disk. Less associated with BCs; attenuation from the ISM dominates.
  
  \vspace{1mm}
  
  \item \emph{Old stars:} Found in the disk and bulge. Those behind the ISM will experience attenuation; those in front will not be attenuated by the ISM.
\end{enumerate}

We are now equipped to test the validity of this model. When looking at separate galaxies on a spaxel-by-spaxel basis, as in Fig.~\ref{fig:gradient_radius_graphs_mass} and Fig.~\ref{fig:gradient_radius_graphs_sSFR}, decreasing radial gradients are found for $A_{V, \: \rm stars}$. Such gradients can reasonably be attributed to the radial density gradient exhibited by the diffuse, dusty ISM of the host galaxies \citep[e.g.][]{Peletier1995TheGalaxies, Boissier2004TheGalaxies, Munoz-Mateos2009RADIALPROPERTIES, Wild2011a, GonzalezDelgado2015TheSequence, Goddard2016SDSS-IVType}.

Ostensibly, a similar explanation might plausibly account for the similar radial gradient exhibited by $A_{V, \: \rm gas}$, since a `screen-like' dust attenuation due to the ISM would affect both the old and the young stellar populations in a similar manner. However, this is not in fact the case; such gradients in the dust attenuation derived from the Balmer decrement cannot be explained by a density gradient in the diffuse ISM alone. Both Fig.~\ref{fig:global_histograms} and Table~\ref{tab:global_ratios} show that the distributions of both the global light-weighted $A_{V, \: \rm stars}$ and $A_{V, \: \rm gas}$ increase as the relative global $\rm sSFR_{H \alpha}$ of the galaxy increases; this behaviour is a natural consequence of a diffuse ISM which is denser for galaxies with higher sSFRs \citep[e.g.][]{DaCunha2008, daCunha2010NewGalaxies, Rowlands2014Herschel-ATLAS:Redshifts, Remy-Ruyer2015LinkingPicture}. However, the ratio of the two attenuation measures is not dependent upon the relative global $\rm sSFR_{H \alpha}$ of the galaxy. This result is in stark contrast to Fig.~\ref{fig:local_histograms}, in which each of the local $A_{V, \: \rm stars}$, the local $A_{V, \: \rm gas}$, and the local $A_{V, \: \rm stars} / A_{V, \: \rm gas}$ all increase as the local $\Sigma_{\rm SFR, \: \rm H \alpha}$ increases. This behaviour suggests that while the radial gradient in $A_{V, \: \rm stars}$ is likely due to a combination of local and global physics (i.e. a diffuse ISM with a radial profile), the seemingly similar gradient in $A_{V, \: \rm gas}$ is instead dominated by purely local physics.

The above results suggest that the concentration of BCs (and associated young stars) increases towards the centre of SF spiral galaxies. Radial gradients have been found in SF galaxies for the dust attenuation derived from the Balmer decrement, which we reiterate is a proxy for the dust attenuation due to the BCs \citep[e.g.][]{Nelson2015Spatially-resolvedZ1.4, Jafariyazani2019SpatiallyQuenching}; SFR per unit physical surface area \citep[e.g.][]{Nelson2015WhereZ1, GonzalezDelgado2016StarGalaxies, Medling2018TheFormation, Sanchez2019Spatially-ResolvedGalaxies}; and the gas mass per unit physical surface area (e.g. \citealp{Bigiel2008TheScales, Colombo2017TheSequence, Sanchez2019Spatially-ResolvedGalaxies}; see also \citealp{Bolatto2017TheCARMA} and \citealp{Lin2017ResolvedMaNGA} who find that the molecular gas content of galaxies peaks at their centres). The fact that all of these closely linked quantities are found to monotonically decrease with distance from the centres of SF galaxies suggests that the same may well be true for the concentration of BCs. The increase in molecular gas at the centres of spiral galaxies, seen both by radio \citep[e.g][]{Bigiel2008TheScales} and IFU \citep[e.g.][]{Lin2017ResolvedMaNGA, Colombo2017TheSequence, Sanchez2019Spatially-ResolvedGalaxies} observations, is especially indicative of centrally concentrated BCs. This is because the molecular gas surface density is related to $\Sigma_{\rm SFR}$ by the Schmidt--Kennicutt Law \citep{Schmidt1959TheFormation., Kennicutt1998TheGalaxies}. We refer the reader to \citet{Bigiel2008TheScales} to see how well $\Sigma_{\rm SFR}$ is linked to the molecular gas content of galaxies.

\begin{figure*}
	\includegraphics[width=\textwidth]{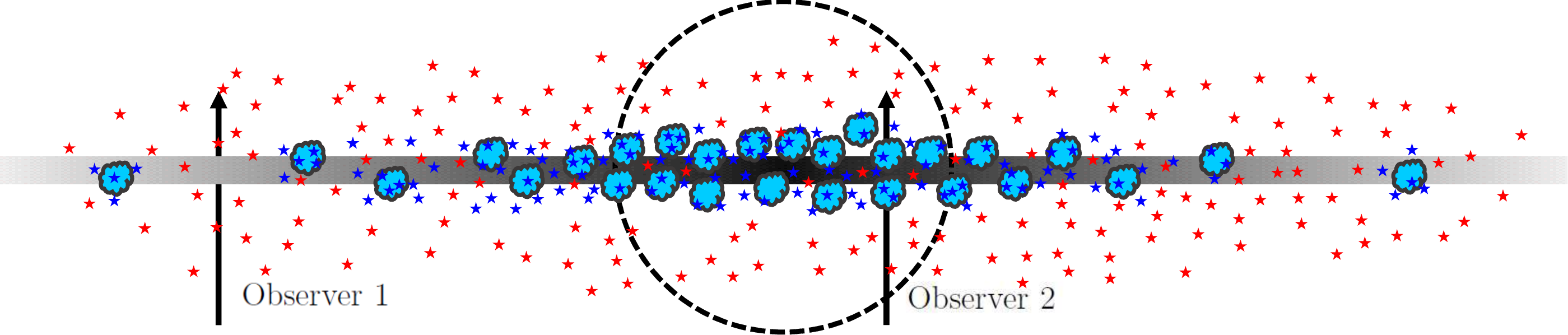}
    \caption{Schematic diagram showing an edge-on view of a SF spiral galaxy. Depicted are four key components: \emph{a diffuse, dusty ISM} with a radial gradient (grey); \emph{dusty, clumpy BCs} surrounded by \textsc{H~ii} regions (light blue clouds); \emph{young stars} either embedded in, or still closely associated with, these BCs (dark blue stars); and \emph{intermediate / old stars} further from their BCs (red stars). No assumptions are made about the bulge of the galaxy (dashed black line). Shown also are the lines of sight of two observers. Observer~1 will see a small amount of dust attenuation affecting the stellar populations, but no dust attenuation due to the clumpy BCs; Observer~2, however, will see high dust attenuation both due to the diffuse ISM and the clumpy BCs. The results in this work suggest that in addition to the presence of a diffuse, dusty ISM, SF spiral galaxies also exhibit a higher concentration of young stars and dusty BCs nearer to their central regions than at their outskirts.}
    \label{fig:dust_geometry}
\end{figure*}

A schematic diagram depicting the dust geometry of SF spiral galaxies described above is shown in Fig.~\ref{fig:dust_geometry}. This dust geometry is similar to that suggested by Fig.~13 of \citet{Wild2011a}, except that in Fig.~\ref{fig:dust_geometry} we explicitly depict the radial dependence of the BCs, since we find direct evidence for such behaviour. The radial gradients seen for $A_{V, \: \rm stars}$ and the $A_{V, \: \rm gas}$ excess in Fig.~\ref{fig:gradient_radius_graphs_mass} and Fig.~\ref{fig:gradient_radius_graphs_sSFR} can be explained by the geometry in Fig.~\ref{fig:dust_geometry}. The line of sight of Observer~1 is far from the galactic centre; therefore, Observer~1 will find low $A_{V, \: \rm stars}$ due to the radial dependence of the ISM, and almost zero $A_{V, \: \rm gas}$ excess due to the absence of BCs along their line of sight. By contrast, Observer~2 has a line of sight close to the galactic centre, and will consequently find higher values for both $A_{V, \: \rm stars}$ and the $A_{V, \: \rm gas}$ excess.

Further evidence for the concentration of clumpy BCs increasing nearer to the centre of SF spiral galaxies can be found in Fig.~\ref{fig:gradient_radius_graphs_mass}, in which the $A_{V, \: \rm gas}$ excess exhibits a radial gradient in all but the lowest-mass galaxies. This behaviour suggests that the radial gradient in the $A_{V, \: \rm gas}$ is produced by a combination of radial gradients in both the diffuse dusty ISM and the clumpy BCs. The radial gradient in the $A_{V, \: \rm gas}$ excess becomes progressively more pronounced for higher-mass galaxies; more massive galaxies, therefore, have clumpier dust distributions and a greater amount of clumpy dust nearer to their centres than less massive galaxies. Fig.~\ref{fig:global_histograms} shows that stellar mass is indeed much more important than relative sSFR for determining the global dust and gas properties of a galaxy; this result is in agreement with the findings of others such as \citet{Garn2010PredictingGalaxy} and \citet{Zahid2012TheGalaxies}. It is stellar mass that is the dominant factor in determining these global properties, since higher stellar masses cause both higher SFRs \citep[e.g][]{Brinchmann2004TheUniverse, Noeske2007StarGalaxies, Peng2010MASSFUNCTION, Whitaker2012THE2.5} and higher gas metallicities \citep[e.g.][]{Tremonti2004TheSDSS, Mannucci2010AGalaxies, Garn2010PredictingGalaxy, Zahid2012TheGalaxies}. Galaxies with higher stellar masses have higher gas metallicities because they are able to retain a greater proportion of their metals \citep[e.g.][]{Tremonti2004TheSDSS}; this in turn leads to increased dust attenuation, since a greater amount of metals are locked into dust grains \citep{Heckman1998TheHigh-Redshift, Boissier2004TheGalaxies}.

In summary, not only do higher-mass galaxies have higher dust attenuations in the diffuse ISM, but they also have even higher dust attenuations due to their BCs. The dust geometry in high-mass galaxies must therefore be clumpier than in low-mass galaxies. Furthermore, a greater concentration of BCs is found nearer to the centres of high-mass galaxies than at their outskirts. This in turn suggests that star formation may be more centrally concentrated in higher-mass SF spiral galaxies.

\section{Conclusions}
\label{sec:Conclusions}

We quantify the dust attenuation for 232 star-forming (SF) MaNGA spirals using two complementary methods. The Balmer decrement is determined across each galaxy's spaxels, and used to measure the dust attenuation in the gas. This quantity is compared with the dust attenuation affecting the stellar populations of each galaxy, ascertained using the full-spectrum stellar population synthesis code \texttt{STARLIGHT} \citep{CidFernandes2005Semi-empiricalMethod}. The results in this work can be summarised as follows:

\begin{enumerate}

  \item The mean global light-weighted dust attenuation affecting the stellar populations ($A_{V, \: \rm stars}$) is consistently lower than the mean global light-weighted dust attenuation derived from the Balmer decrement ($A_{V, \: \rm gas}$). The mean $A_{V, \: \rm stars}$ is not strongly affected by the stellar mass of the galaxy, but the mean $A_{V, \: \rm gas}$ increases for increasing stellar mass. Both the mean values of $A_{V, \: \rm stars}$ and $A_{V, \: \rm gas}$ decrease as the global $\rm H \alpha$ derived specific star formation rate ($\rm sSFR_{H \alpha}$) relative to the star-forming main sequence (SFMS) of the galaxy decreases. The ratio of these two attenuation measures decreases for increasing stellar mass, but is independent of the global $\rm sSFR_{H \alpha}$ relative to the SFMS of the galaxy. The calculated ratios are comparable to, but consistently lower than, those found by \citet{Calzetti2000TheGalaxies}.
    
  \vspace{1mm}
  
  \item No difference is found for any of the local dust attenuation quantities analysed in this work between the spiral arm and inter-arm regions of the galaxies. This null result can largely be attributed to the relatively poor spatial resolution of MaNGA, which is unable to resolve narrow dust lanes within the spiral arms. As such, this investigation would be worth revisiting with higher resolution data.
  
  \vspace{1mm}
  
  \item Both the local $A_{V, \: \rm stars}$ and $A_{V, \: \rm gas}$ decrease with galactocentric radius, in agreement with the findings of \citet{GonzalezDelgado2015TheSequence, Goddard2016SDSS-IVType} ($A_{V, \: \rm stars}$) and \citet{Nelson2015Spatially-resolvedZ1.4, Jafariyazani2019SpatiallyQuenching} ($A_{V, \: \rm gas}$). By contrast, the local $A_{V, \: \rm stars} / A_{V, \: \rm gas}$ ratio does not vary between the centre of SF spiral galaxies and their outer regions (out to $1.5 \ R_{\rm e}$). The exception is the lowest-mass galaxies, which exhibit some radial gradient for this ratio due to the absence of such a gradient in their local $A_{V, \: \rm gas}$.
  
  \vspace{1mm}
  
  \item We introduce a new dust attenuation quantity, which we coin the $A_{V, \: \rm gas}$ excess (defined as $A_{V, \: \rm gas} - A_{V, \: \rm stars}$). This quantity, unlike the $A_{V, \: \rm stars} / A_{V, \: \rm gas}$ ratio, is physically meaningful, providing an estimate for the dust attenuation due the birth clouds (BCs) alone. A radial gradient is also exhibited by the $A_{V, \: \rm gas}$ excess, thus demonstrating that the radial gradient in the local $A_{V, \: \rm gas}$ cannot be attributed to a gradient in the diffuse ISM alone; the concentration of clumpy dust must also increase nearer to the galactic centre.
  
  \vspace{1mm}
  
  \item We utilise the spatial resolution of MaNGA to analyse the the local values of each of $A_{V, \: \rm stars}$, $A_{V, \: \rm gas}$, $A_{V, \: \rm stars} / A_{V, \: \rm gas}$, and $A_{V, \: \rm gas}$ excess. Each of these quantities increases as the local star formation rate per unit physical surface area increases. These results suggest that variations in dust attenuation properties are likely driven predominantly by local physics, rather than the global properties of the galaxies.
  
  \vspace{1mm}
  
  \item The results in this work are consistent with the model that stars are born in giant molecular BCs in which gas and very clumpy dust are highly concentrated \citep{LonsdalePersson1987OnFluxes, Charlot2000AGalaxies, Calzetti2000TheGalaxies}. However, this work demonstrates that the dust geometry in high-mass galaxies is clumpier than in low-mass galaxies. Furthermore, SF spiral galaxies are also shown to exhibit a much higher concentration of both young stars and BCs nearer to their central regions than at their outskirts.

\end{enumerate}

\section*{Acknowledgements}

We thank the anonymous referee for their very positive and useful comments and suggestions included in their review.

Funding for the Sloan Digital Sky Survey IV has been provided by the Alfred P. Sloan Foundation, the U.S. Department of Energy Office of Science, and the Participating Institutions. SDSS acknowledges support and resources from the Center for High-Performance Computing at the University of Utah. The SDSS website is \url{www.sdss.org}.

SDSS is managed by the Astrophysical Research Consortium for the Participating Institutions of the SDSS Collaboration including the Brazilian Participation Group, the Carnegie Institution for Science, Carnegie Mellon University, the Chilean Participation Group, the French Participation Group, Harvard-Smithsonian Center for Astrophysics, Instituto de Astrof{\'i}sica de Canarias, The Johns Hopkins University, Kavli Institute for the Physics and Mathematics of the Universe (IPMU) / University of Tokyo, the Korean Participation Group, Lawrence Berkeley National Laboratory, Leibniz Institut f{\"u}r Astrophysik Potsdam (AIP), Max-Planck-Institut f{\"u}r Astronomie (MPIA Heidelberg), Max-Planck-Institut f{\"u}r Astrophysik (MPA Garching), Max-Planck-Institut f{\"u}r Extraterrestrische Physik (MPE), National Astronomical Observatories of China, New Mexico State University, New York University, University of Notre Dame, Observat{\'o}rio Nacional / MCTI, The Ohio State University, Pennsylvania State University, Shanghai Astronomical Observatory, United Kingdom Participation Group, Universidad Nacional Aut{\'o}noma de M{\'e}xico, University of Arizona, University of Colorado Boulder, University of Oxford, University of Portsmouth, University of Utah, University of Virginia, University of Washington, University of Wisconsin, Vanderbilt University, and Yale University.

This publication uses data generated via the \url{zooniverse.org} platform, development of which is funded by generous support, including a Global Impact Award from Google, and by a grant from the Alfred P. Sloan Foundation. This publication has been made possible by the participation of almost 6000 volunteers in the Galaxy Zoo:3D project on \url{zooniverse.org}.

This research made use of \texttt{Astropy}, a community-developed core \texttt{Python} (\url{https://www.python.org}) package for Astronomy \citep{Robitaille2013Astropy:Astronomy}; \texttt{Matplotlib} \citep{Hunter2007}; \texttt{NumPy} \citep{VanderWalt2011}; \texttt{SciPy} \citep{Virtanen2019SciPyPython}; and \texttt{TOPCAT} \citep{Taylor2005}.

M.B. acknowledges the FONDECYT regular grant 1170618. This work was supported by the Science and Technology Facilities Council.




\bibliographystyle{mnras}
\bibliography{main.bib} 




\bsp	
\label{lastpage}
\end{document}